\documentstyle[jer]{article}
\def\href#1#2{#2}
\begin{document}
\newcommand{\bfi}{{\bf B}} \newcommand{\efi}{{\bf E}}
\newcommand{\lel}{{\lambda_e^{\!\!\!\!-}}}
\newcommand{\me}{m_e}
\newcommand{\mcs}{{m_e c^2}}
\title{Almost Analytic Models of Ultramagnetized Neutron Star Envelopes}
\author{Jeremy S. Heyl\thanks{Current address: Theoretical Astrophysics, mail code 130-33, California Institute of Technology, Pasadena CA 91125, USA} \and Lars Hernquist\thanks{Presidential Faculty
Fellow}}
\maketitle
\begin{center}
Lick Observatory,
University of California, Santa Cruz, California 95064, USA
\end{center}
\begin{abstract}

Recent ROSAT measurements show that the x-ray emission from isolated
neutron stars is modulated at the stellar rotation period.  To
interpret these measurements, one needs precise calculations of the
heat transfer through the thin insulating envelopes of neutron stars.
We present nearly analytic models of the thermal structure of the
envelopes of ultramagnetized neutron stars.  Specifically, we examine
the limit in which only the ground Landau level is filled.  We use the
models to estimate the amplitude of modulation expected from
non-uniformities in the surface temperatures of strongly magnetized
neutron stars.  In addition, we estimate cooling rates for stars with
fields $B \sim 10^{15}-10^{16}$~G which are relevant to models that
invoke ``magnetars'' to account for soft $\gamma$-ray emission from
some repeating sources.

\end{abstract}

\section{Introduction}

Since the launch of the ROSAT satellite, our knowledge of isolated
neutron stars has expanded into new realms.  Before ROSAT, neutron
stars were somewhat unique among astronomical objects.  Although they
had been observed over a range of energies from radio to
ultra-high-energy gamma rays, and had been evoked to power a variety
of astrophysical objects from pulsars to soft X-ray repeaters and
gamma-ray bursts, one could not argue unequivocally that a single
photon from the surface of a neutron star had ever been detected.

For the first time, we have direct evidence for radiation from the
surfaces of neutron stars.  More than a dozen such sources have been
detected by ROSAT (\eg\ \cite{Ogel95}), and more than ten have been
fitted with spectra.  The spectra divide the objects into two classes:
1) objects with hard spectra whose X-ray emission is best attributed to
the magnetosphere, and 2) neutron stars whose soft flux is
well-described by a blackbody spectrum.  \tabref{psrlist} lists several
of those that fall into the second group with pertinent references.
\figcomment{\bt
\caption{Several pulsars with observed surface blackbody emission}
\label{tab:psrlist}
\begin{tabular}{ll}
Pulsar & References \\ \hline
PSR J0437-4715        & \cite{Beck93} \\
PSR 0630+18 (Geminga) & \cite{Halp92,Halp93} \\
                      & \cite{Halp97} \\
PSR 0656+14           & \cite{Finl92}, \\
                      & \cite{Ande93,Grei96}, \\
                      & \cite{Poss96} \\
PSR 0833-45 (Vela)    & \cite{Ogel93b} \\
PSR 1055-52           & \cite{Ogel93a,Grei96} \\
PSR 1929+10           & \cite{Yanc94}
\end{tabular}
\et}

The X-ray spectra of these objects consist of a soft blackbody and
hard power-law component.  Generally, the surface temperature inferred
from modeling the blackbody is approximately what one would expect
for a cooling neutron star at the characteristic age of the pulsar.
If this interpretation is correct, the observations provide a direct
probe of the structure of cooling neutron stars.  Furthermore, the
observations show that the thermal radiation is modulated at the
rotation period of the pulsars.  A strong magnetic field can modulate
the thermal flux by causing the heat conduction in the outer layers of
the star to be anisotropic.  Both to translate these observations
into constraints on the structure of neutron stars and to understand
modulation of the radiation, we must have a detailed understanding of
the insulating layers of the neutron-star crust, \ie the envelope. 

We proceed in the spirit of \jcite{Gudm82} and concentrate our
analysis on the thin region, the envelope, which insulates the bulk of
the neutron star.  The envelope is customarily defined to extend from
zero density to $\rho \sim 10^{10}$~g/cm$^3$, and its thickness
($h_E$) is of the order of tens of meters, very small compared to the
radius of the star, $R\sim 10$ km.  By limiting the analysis, we focus
on how various physical processes affect the thermal structure of the
envelope and the relationship between the core temperature and the
flux emitted at the surface.  An alternative point of view is to
combine the envelope calculation with an estimate of the cooling rate
due to neutrinos and the total heat capacity of the neutron star,
yielding theoretical cooling curves (\eg\
\cite{Tsur72,Nomo81,Glen80,VanR81})

Several authors have made much progress in understanding the properties
of neutron star envelopes with and without magnetic fields.
\jcite{Gudm82} numerically calculate the thermal structure for
unmagnetized envelopes, and \jcite{Hern84b} present analytic models for
the $B=0$ case.  \jcite{Tsur79}, \jcite{Glen80}, \jcite{VanR81} and
\jcite{Urpi80} calculate the luminosity observed at infinity as a
function of the core temperature for several magnetic field strengths
less than $10^{14}$~G, including the zero-field case.  \jcite{Hern85}
calculates the thermal structure of envelopes for $B \le\ 10^{14}$~G for
transport along the field, using the electron conductivities of
\jcite{Hern84a} which account for the quantization of electron
energies in the magnetic field in a relativistic framework.  We will
use these conductivities in the present work; therefore,
\jcite{Hern85} provides a natural benchmark.

\jcite{VanR88} builds upon the \jcite{Hern85} results by exploring
various assumptions concerning the properties of the envelope at low
densities and calculating profiles for many field strengths
($B<10^{14}$~G) and core temperatures.  Again, these calculations are
limited to conduction along the field.  \jcite{Scha90a}, using the
electron conductivities calculated in \jcite{Scha88}, calculates the
thermal structure in two dimensions for $B \lsim 10^{11}$~G.  Above
a field strength of $10^{12}$~G, the calculations are not considered
reliable.  Finally, \jcite{Shib95} present the temperature distribution
as a function of magnetic colatitude for $B=10^{12}$~G from a
numerical solution to the two-dimensional thermal structure equation in
a plane-parallel approximation.  

The current work complements the previous ones by extending the results
to stronger field strengths ($10^{14} \rmmat{ G} \le\ B \le\ 10^{16}$~G)
in a semi-analytical fashion.  We apply the approach of
\jcite{Hern84b} in the limit of a strongly magnetized envelope, and
then justify and use the plane-parallel approximation to solve the
two-dimensional structure equation.  We derive separable thermal
structure equations in the high and low temperature limits for both
liquid and solid material.  We calculate the thermal structure in
terms of simple (although analytically intractable) integrals.

The plane-parallel approximation has the second important advantage
that the detailed field configuration separates from the thermal
structure problem.  Assuming that it is correct, we can synthesize the
results for any field distribution $B(\theta,\phi)$ as long as $B$ is
not too inhomogeneous on the scale of the envelope thickness (\ie
$|B/\nabla B| \gg\ h_E$).

We find that the emission from a given surface element is a simple
function of the location of the element.  Using this functional form, we
derive light curves and time-dependent spectra including general
relativistic effects.  Although we closely follow the formalism of
\jcite{Page95}, we calculate the two-dimensional thermal structure of
the envelope and present results for several field strengths and fluxes.
We internally verify and justify the geometric simplification used to
translate our results into observables.

\section{Preliminaries}

In extremely intense magnetic fields, the Landau energy
($\hbar\omega_B$) of an electron will typically exceed its thermal
energy.  In these strong fields, the quantization of the electron energy
determines the structure of the electron phase space and must be taken
into account in calculating the thermodynamics of the electron gas. 

In what follows, we will use the dimensionless units
\ba
\beta &=&  \frac{\hbar \omega_B}{\mcs} = \frac{\hbar |e|}{\me^2 c^3} B
\approx \frac{B}{4.4 \times 10^{13} \rmmat { G}}, \\
\label{eq:betadef}
\tau &=& \frac{k T}{\mcs} \approx \frac{T}{5.9 \times 10^9 \rmmat{ K}}
\rmmat{~and~} \zeta = \frac{\mu}{\mcs} 
\label{eq:zetataudef}
\ea
where $\mu$ is the chemical
potential of the electron gas including the electron rest mass.
\jcite{Heyl98thesis}, \jcite{Hern85}, and \jcite{Hern84a} outline the
techniques for calculating the thermodynamic properties of a
magnetized electron gas, and we adopt their methodology for the 
remainder of the paper.
\newcommand{\kzz}{\kappa_{zz}}
\newcommand{\kyy}{\kappa_{yy}}

\section{The Low-Temperature, Strong-Field Regime}

We are specifically interested in the low temperature limit ($\tau \ll \zeta-1$)
and the regime in which only one Landau level is filled
($\zeta < \sqrt{2\beta+1}$).  For neutron stars with $\beta \gsim 1$,
this limit applies to the regions that most effectively insulate the
isothermal core of the star.  We will use the results of 
\jcite{Yako84} and \jcite{Hern84a} to calculate the thermal conduction
in the liquid and solid phases.

\subsection{Degenerate Structure Equations}

If we assume that the pressure is supplied by the electrons alone, the
general relativistic equations of thermal structure in the
plane-parallel approximation assume the simple form
\ba
\dd{T}{\mu} &=& \frac{F}{g_s} \frac{Y_e}{m_u} \frac{1}{\kappa}
 \left ( 1 - \frac{F}{g_s}
\frac{S_e}{\rho \kappa} \right )^{-1}
\label{eq:dTdmu} \\
\dd{\mu}{z} &=& \frac{m_u}{Y_e} g_s \left ( 1 - \frac{F}{g_s}
\frac{S_e}{\rho \kappa} \right )
\label{eq:dmudz}
\ea
where we have neglected the thickness of the envelope ($h_E\sim 100$
m) relative to the stellar radius ($R$).  Here, $m_u$ is the atomic
mass unit, $Z$ and $A$ are the mean atomic number and mean atomic mass
of the material, $\rho$ is the density of the matter, $S_e$ is the
entropy of the electron gas per unit volume, $\kappa$ is the thermal
conductivity, and $F$ and $g_s$ are the flux and the acceleration of
gravity as measured at the surface, respectively.  For completely
ionized material, $Y_e$ is given by the product of $Z/A$ and the
ionized fraction.

In the absence of a magnetic
field, this plane-parallel approximation introduces errors of the
order $R_s h_E/R^2 \approx 0.6 \%$ where $R_s=2GM/c^2$
(\cite{Gudm82}).  To understand the potential errors of the
plane-parallel treatment in the presence of a magnetic field, we
compare the results of \jcite{Shib95} with those of \jcite{Scha90a}.
Although \jcite{Shib95} use a one-dimensional approach, their results
agree with those of the two-dimensional calculations by \jcite{Scha90a}
for several surface temperatures and a magnetic field of $10^{12}$ G.
In stronger magnetic fields, conduction perpendicular to the magnetic
field is even less important, and we expect the one-dimensional
method to be even more accurate.

To estimate the errors in using the plane-parallel treatment in the
presence of the magnetic field, we examine the thermal structure
equation in two dimensions (\cite{Scha90a})
\def\expL{e^{-\Lambda_s}}
\ba
\bfm{\nabla \cdot F} &=& -\frac{\expL}{r^2} \pp{}{r} \left [ r^2
\kappa_{11} \expL \pp{T}{r} + r \kappa_{12} \pp{T}{\theta} \right ]
\\ \nonumber
& & ~~~ -\frac{1}{r \sin \theta} \pp{}{\theta} \left [ \sin \theta \left (
\kappa_{21} \expL \pp{T}{r} + \frac{\kappa_{22}}{r}
\pp{T}{\theta} \right ) \right ] = 0,
\ea
where $\theta$ is an angle along the surface of star, specifically the
magnetic colatitude,
\be
\expL = \sqrt{1-\frac{R_s}{R}},
\label{eq:expLdef}
\ee
and the $\kappa_{ij}$ are the components of the thermal conduction tensor,
where 1 denotes the radial direction and 2 denotes the tangential
direction.  The components of $\bfm{\kappa}$ are found by
rotating the tensor calculated by \jcite{Hern84a} and \jcite{Yako84}
so that the $z$-direction locally coincides with the radial direction.  
This gives
\ba
\kappa_{11} &=& \kyy \sin^2 \psi + \kzz \cos^2 \psi, \\
\kappa_{22} &=& \kyy \cos^2 \psi + \kzz \sin^2 \psi, \\
\kappa_{12} = \kappa_{21} &=&  (\kyy-\kzz) \sin \psi \cos \psi,
\ea
where $\psi$ is the angle between the local field direction and the
radial direction, and $\kyy$ and $\kzz$ respectively are the components
of the heat conduction tensor perpendicular and parallel to the
direction of the magnetic field.

For a uniformly magnetized neutron star
$\psi=\theta$; for a dipole field, $\cot \psi=2 \cot \theta$
(\cite{Gree83}), or more conveniently
\be
\cos^2 \psi = \frac{4 \cos^2 \theta}{3 \cos^2 \theta + 1}.
\label{eq:cos2psi}
\ee

If we assume that the components of the thermal conduction
matrix ($\bfm{\kappa}$) are of the same order and take
the maximum temperature gradient to be
$T_c-T_\rmscr{eff} \sim T_c$ radially over the thickness of the
envelope, $h_E$, or tangentially over one radian, we obtain
\be
\pp{T}{\theta} \sim T_c \ll \expL r \pp{T}{r} \sim
\expL R \frac{T_c}{h_E} \sim 10^2 T_c
\ee
where $T_c$ is the core temperature.  We find that neglecting
derivatives with respect to angle does not dramatically increase the
error relative to the unmagnetized plane-parallel case.

However, this argument does not apply for $\theta$ close to $\pi/2$
(\ie where the magnetic field lines are parallel to the surface for a
uniform or dipole field).  Here,
\be
\kappa_{22} = \kappa_{zz} \gg \kappa_{11} = \kappa_{yy}.
\ee
If we reexamine the error analysis for $\psi\approx\pi/2$ we
find that relevant quantities to compare are
\be
\kappa_{22} \frac{\partial^2 T}{\partial \theta^2} \rmmat{~and~}
\kappa_{11} r^2 e^{-2\Lambda_s} \frac{\partial^2 T}{\partial r^2}.
\ee
The tangential transport will exceed the radial transport if
\be
\cos^2 \psi <  e^{2\Lambda_s}\frac{h_E^2}{R^2} - \frac{\kappa_{yy}}{\kappa_{zz}}
\ee
Comparing these values we find that if $\kappa_{zz} \gsim 10^4
\kappa_{yy}$, the one-dimensional treatment will break down near
$\psi=\pi/2$; otherwise, the plane-parallel treatment is adequate even
at $\psi=\pi/2$. 

For regions where the magnetic field lines are not nearly parallel to
the surface, the plane-parallel approximation works well; consequently
even for an arbitrary field geometry, because the envelope is thin, we
ignore $\partial/\partial\theta$ terms in the structure equation
compared to $r \partial/\partial r$ terms and focus
on radial heat flow.  With these assumptions, we have (\cite{Scha90a})
\be
\kappa = \kappa_{11} = \kzz \cos^2 \psi + \kyy \sin^2 \psi.
\label{eq:kappageom}
\ee

In the low-temperature limit, we obtain the dimensionless equation
\be
\dd{\tau}{\zeta} = \left ( Y_e \frac{F}{m_u g_s} \frac{\lel^2}{c} \right )
\left ( \frac{c k}{\lel^2} \frac{1}{\kappa} \right )
\label{eq:dtaudzeta}
\ee
where $\lel$ is the electron Compton wavelength and
the dimensionless flux is given by
\be
\left ( Y_e \frac{F}{m_u g_s} \frac{\lel^2}{c} \right ) = 7.83 \times 10^{-3}
\frac{Z_{26}}{A_{56}} \frac{T_\rmscr{eff,6}^4}{g_{s,14}},
\label{eq:fluxconstant}
\ee
where $Z_{26}=Z/26$, $A_{56}=A/56$, $T_\rmscr{eff,6}=T_\rmscr{eff}/10^6$ K,
and $g_{s,14}=g_s/10^{14}$ cm/s$^2$.  $T_\rmscr{eff}$ is the effective
blackbody temperature of the neutron star photosphere again as measured
at the surface, which we take to be located at an optical depth of 2/3. 

For \eqref{dtaudzeta} to be separable for an arbitrary geometry, $\kzz$
and $\kyy$ must depend on $\tau$ in the same fashion.  For electron-ion
scattering this is the case, so we can hope to find a simple analytic
solution to the structure equation.  Unfortunately, since the
cross-section for electrons to scatter off of phonons depends explicitly
on temperature, in the solid state the structure
equation is separable only where the field is either purely
radial or tangential.  

In the liquid state, we obtain the following structure equation,
\be
\tau \dd{\tau}{\zeta} = \left ( Y_e \frac{F}{m_u g_s} \frac{\lel^2}{c} \right )
\left [ \cos^2 \psi \frac{\pi}{3} \frac{\beta^2}{Z \alpha^2} 
\frac{\phi_{ei}(\zeta;\beta)}{\sqrt{\zeta^2-1}}
+ \frac{\sin^2 \psi}{12\pi} \frac{Z \alpha^2}{\beta} 
Q_{ei}(\zeta;\beta) \sqrt{\zeta^2-1}
 \right ]^{-1}
\label{eq:dtaudzetaliquid}
\ee
where $\alpha$ is the fine-structure constant.
In the solid state for electron-phonon scattering, we obtain
\ba
\dd{\tau}{\zeta} &=&
\left ( Y_e \frac{F}{m_u g_s} \frac{\lel^2}{c} \right )
\left [ \frac{1}{3} \frac{\beta^2}{\alpha u_{-2}} \phi_{ep}(\zeta;\beta)
\right ]^{-1}\!\!,~
 \psi=0
\label{eq:dtaudzetasolid0}
\\
\tau^2 \dd{\tau}{\zeta} &=&
\left ( Y_e \frac{F}{m_u g_s} \frac{\lel^2}{c} \right )
\left [ \frac{1}{12} \frac{\alpha u_{-2}}{\beta} Q_{ep}(\zeta;\beta)
\right ]^{-1}\!\!,~
 \psi=\frac{\pi}{2}
\label{eq:dtaudzetasolid90}
\ea
The functions $\phi$ and $Q$ are defined and calculated in
\jcite{Yako84} and \jcite{Hern84a} and we take $u_{-2} \approx 13$
(\cite{Yako80,Pote96b}) for a body-centered cubic lattice.

\section{The High Temperature Regime}

In the nondegenerate regime we assume that most of the heat is transported by
photons and that free-free absorption provides the opacity.  We take the
unmagnetized thermal conductivity to be of the Kramer's form (\cite{Sila80}),
\be
\kappa^{(F)} =
\kappa_0 \frac{T^{13/2}}{\rho^2} \eta_{ff}(b,\psi)
\label{eq:kappaKramer}
\ee
where
\ba
\kappa_0 &=& \frac{1}{2.947} \frac{c_7}{\pi\sqrt{2 \pi}}
\frac{\sigma c k^{7/2} m_u^2 m_e^{3/2}}{e^6 \hbar^2}
\frac{A^2}{Z^3} \\
&=& \frac{16 \sigma}{3} m_u \frac{196.5}{24.59} \frac{A^2}{Z^3}
\frac{\rmmat{g}}{\rmmat{cm}^5\rmmat{K}^{7/2}},
\ea
$c_7=316.8$, $\sigma$ is the Stefan-Boltzmann constant and 
$b\equiv \beta/\tau$. The factor of 2.947 scales the results of
 \jcite{Sila80}
to agree with the results of \jcite{Cox68} (for discussion see
\cite{Hern85}).

We parameterize the effects of the magnetic field by the anisotropy
factor for free-free absorption ($\eta_{ff}$).  Absorption dominates
the opacity through the nondegenerate portion of the envelope
(\cite{Pavl77,Sila80}). We use the analytic results of \jcite{Pavl77}
to extrapolate beyond the tabulated values in \jcite{Sila80}; \ie for
$b>1000$.  For $b<1000$ we use the results of \jcite{Sila80}.

\subsection{Non-degenerate Structure Equation}

We follow the method of \jcite{Hern84b}, but we do not require that
the conductivity be a power law in $\rho$ and $T$.

In the non-degenerate regime, the thermal structure equation of the
envelope is
\be
\dd{T}{P} = \frac{F}{g_s} \frac{1}{\rho \kappa}
\label{eq:ndstructure}
\ee
and we consider an unmagnetized conductivity which is a power law as
\eqref{kappaKramer}
\be
\kappa = \kappa_0 \frac{T^\delta}{\rho^{\alpha}}.
\label{eq:kappapower}
\ee
Even in an intense magnetic field, in the nondegenerate regime,
the pressure is given by the ideal gas law (\cite{Blan82})
\be
P = \frac{Y_e}{m_u} \rho k T.
\label{eq:ndstate}
\ee
We combine Equations~\ref{eq:ndstructure}--\ref{eq:ndstate} with
Equation~\ref{eq:kappaKramer}, yielding
\be
\dd{T}{P} = \frac{F}{g_s} \frac{1}{\kappa_0} \left ( \frac{m_u}{Y_e k}
\right )^{\alpha-1}
 \frac{P^{\alpha-1}}{\eta_{ff}(b,\psi) T^{\alpha+\delta-1}}.
\ee
As for the structure equations in the degenerate limit, this equation is
separable, yielding $\rho(T)$.  Because $\eta_{ff}$ depends on $T$
through the argument $b$, the relation between $T$ and $P$ need not be a
power law as in the unmagnetized case.  In the limit that $\eta_{ff}=1$,
the result of \jcite{Hern84b} obtains. 

More generally, if we take, $\eta_{ff}$ to be a power law
$\propto b^{-2}$ (\eg\ \cite{Tsur72}) which is
approximately true for $b \rarrow \infty$, we can immediately use the
results of \jcite{Hern84b} to obtain that the conductivity is constant
along a solution through the nondegenerate envelope:
\ba
\kappa &=& \frac{\alpha+\delta-2}{\alpha} \frac{F}{g_s} \frac{Y_e
k}{m_u} \\
&=& 
7.07 \times 10^{13} \frac{Z_{26}}{A_{56}}
\frac{T_\rmscr{eff,6}^4}{g_{s,14}}
\frac{\rmmat{erg}}{\rmmat{K cm s}} 
\label{eq:kappasolnnd}
\ea
for $\alpha=2$ and $\delta=6.5$ as in \eqref{kappaKramer}. 
If we equate this result with our assumed conductivity and
solve for $T$,
we find that the solution follows
\be
T = \left ( \frac{\kappa}{a_\beta \kappa_0} \right )^{1/(\delta-2)}
\rho^{\alpha/(\delta-2)} = 10^6 \rmmat{K} 
\left ( \frac{\rho}{\rho_{T_6}} \right )^{4/9}
\label{eq:Trhond}
\ee
where
\ba
a_\beta &=& \frac{2}{3} q \left ( \frac{ \beta \mcs }{k} \right )^2
\frac{1}{\ln b_\rmscr{Typical}} \\
\rho_{T_6} &=& 71.3 \rmmat{ g cm}^{-3}
\frac{\beta}{\sqrt{\ln b_\rmscr{Typical}}}  
A_{56}^{3/2} Z_{26}^{-2}
T_\rmscr{eff,6}^{-2}
g_{s,14}^{1/2},
\ea
and $b_\rmscr{Typical}$ is a typical value of $\beta/\tau$ in the envelope,
$b_\rmscr{Typical} \approx 6 \times 10^3 \beta$.  One should note that for
free-free scattering in the weak-field limit, $T \propto \rho^{4/13}$.

With \eqref{Trhond}, we can calculate the density at the onset of
degeneracy.  We will assume that at the onset of degeneracy the
electron density is given by fully degenerate expression and that
$\tau \approx \zeta - 1$.  This yields
\be
\rho_\rmscr{ND/D} = 3.92 \times 10^5 \rmmat{ g cm}^{-3} \beta (
\ln b_\rmscr{Typical} )^{1/7} 
A_{56}^{6/7} Z_{26}^{-5/7} T_\rmscr{eff,6}^{4/7}
g_{s,14}^{-1/7}.
\ee

In principle, electron scattering could also play a role in the 
nondegenerate regime.  To
simplify the calculation, we neglect its contribution and verify that
it is indeed negligible.
Using \eqref{Trhond}, we find that
the ratio of free-free to electron scattering opacity 
along a solution is given by
\be
\frac{\tilde{\kappa}^{(F)}_{\beta=0}}{\tilde{\kappa}^{(T)}_{\beta=0}} 
= 1.74 \times 10^6 
\rho_0^{-5/9}
\left ( \frac{\beta}{\sqrt{\ln b_\rmscr{Typical}}} \right )^{14/9} \nonumber \\
A_{56}^{4/3} Z_{26}^{-10/9}
T_\rmscr{eff,6}^{-28/9}
g_{s,14}^{7/9},
\ee
where the electron scattering opacity 
${\tilde{\kappa}^{(T)}_{\beta=0}}$ is given by \jcite{Sila80}.
Since this ratio increases with decreasing density, we only need to
estimate it at the maximum density for the
solution, \ie the density at the onset of degeneracy
\be
\left .
\frac{\tilde{\kappa}^{(F)}_{\beta=0}}{\tilde{\kappa}^{(T)}_{\beta=0}}
\right |_\rmscr{ND/D}
= 1.36 \times 10^3
\beta (\ln b_\rmscr{Typical})^{-6/7} 
A_{56}^{6/7} Z_{26}^{-5/7}
T_\rmscr{eff,6}^{-24/7}
g_{s,14}^{6/7}.
\ee
For $B=10^{14}$~G this ratio is greater than one for
$T_\rmscr{eff}<5.9 \times 10^6$ K, which is larger than the effective
temperatures considered here.  Furthermore, this is a conservative
estimate of this ratio because generally we cut off the nondegenerate
solution where degenerate electrons begin to dominate the heat
conduction.  In unmagnetized envelopes this occurs where the gas is
mildly degenerate (\cite{Hern84b}).  We find that this is also the
case for strongly magnetized envelopes.

\section{Calculations}

\subsection{Strategy}

We have found that the heat transfer equation is not solvable
analytically, but it is separable in several cases.  For the liquid
and degenerate region of the envelope the solution may be calculated
once for each field strength and geometry, and scaled to reflect the
magnitude of the heat flux and shifted to fit the temperature at the
low density edge of the region.  We can apply this same strategy in
the solid state only for the case of a purely radial or azimuthal
field.  Otherwise, for the solid region, the temperature as a function
of chemical potential will depend on the flux and boundary conditions
in a more complicated way; consequently, the solution must be
recalculated for each value of the flux.  To match the solutions
across the liquid-solid phase transition and the
degenerate-nondegenerate interface, we follow the approach of
\jcite{Hern84b}.  At the outer boundary, we use the radiative zero
solution.

\subsection{Results for the separable structure equations}

As described earlier, in the low-temperature limit when only one
Landau level is filled the structure equation is simple and
may be integrated for a given field strength and geometry and
the boundary conditions and the dependence on the flux may be
satisfied after the numerical solution is obtained.

Given these numerical results, it is straightforward to calculate the
core temperature for a given surface temperature and flux.  However,
before fixing the boundary conditions, we can note several general
features of the results.  First, for transport along the magnetic field,
the envelope becomes nearly isothermal at $\zeta-1 \sim 0.1$ regardless of
the magnetic field strength.   However, for transport perpendicular to
the field, the temperature rises steadily throughout the range of
applicability of this formalism.  

\paragraph{Parallel Transport}

\figcomment{\begin{figure}
\plottwo{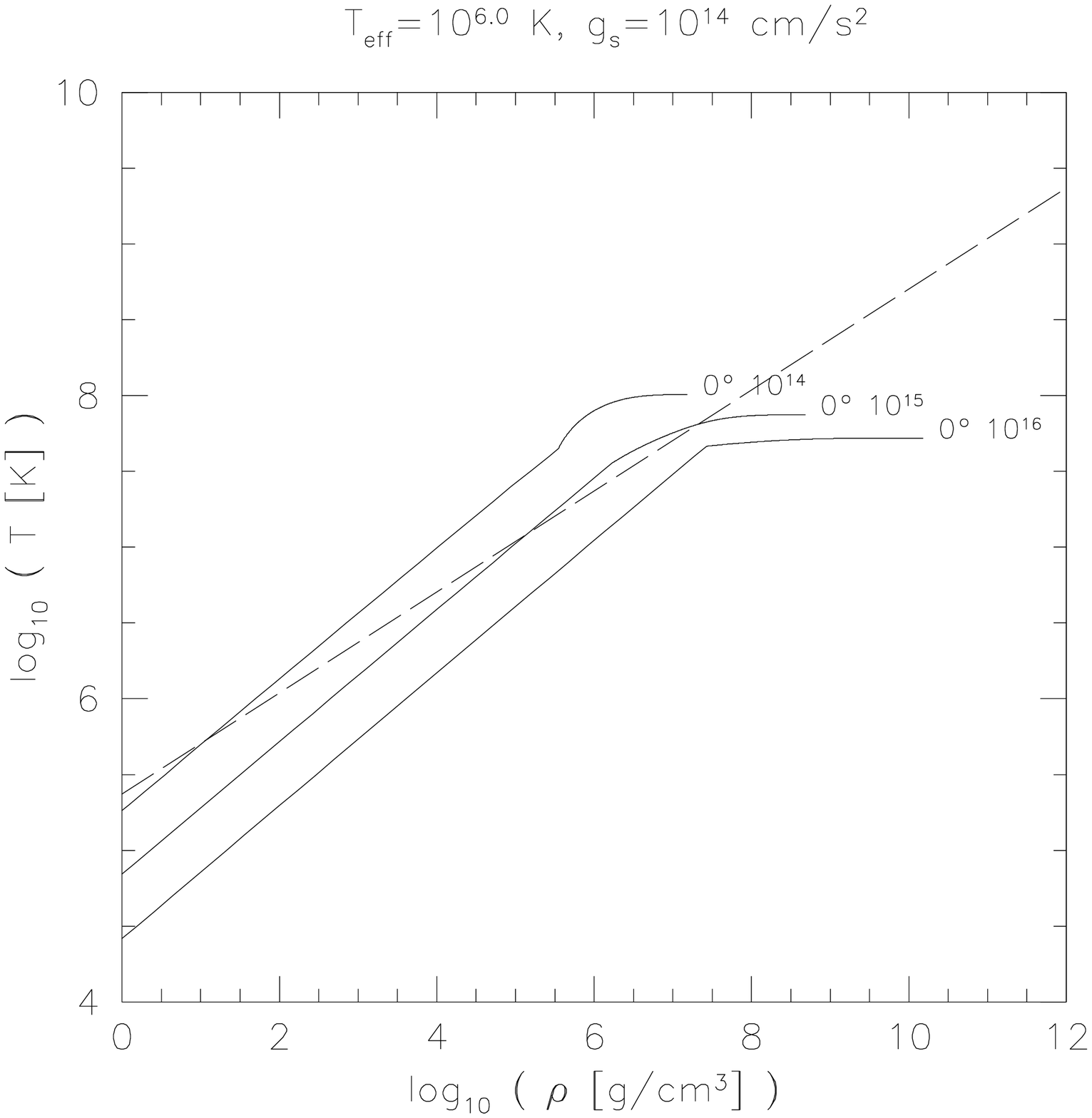}{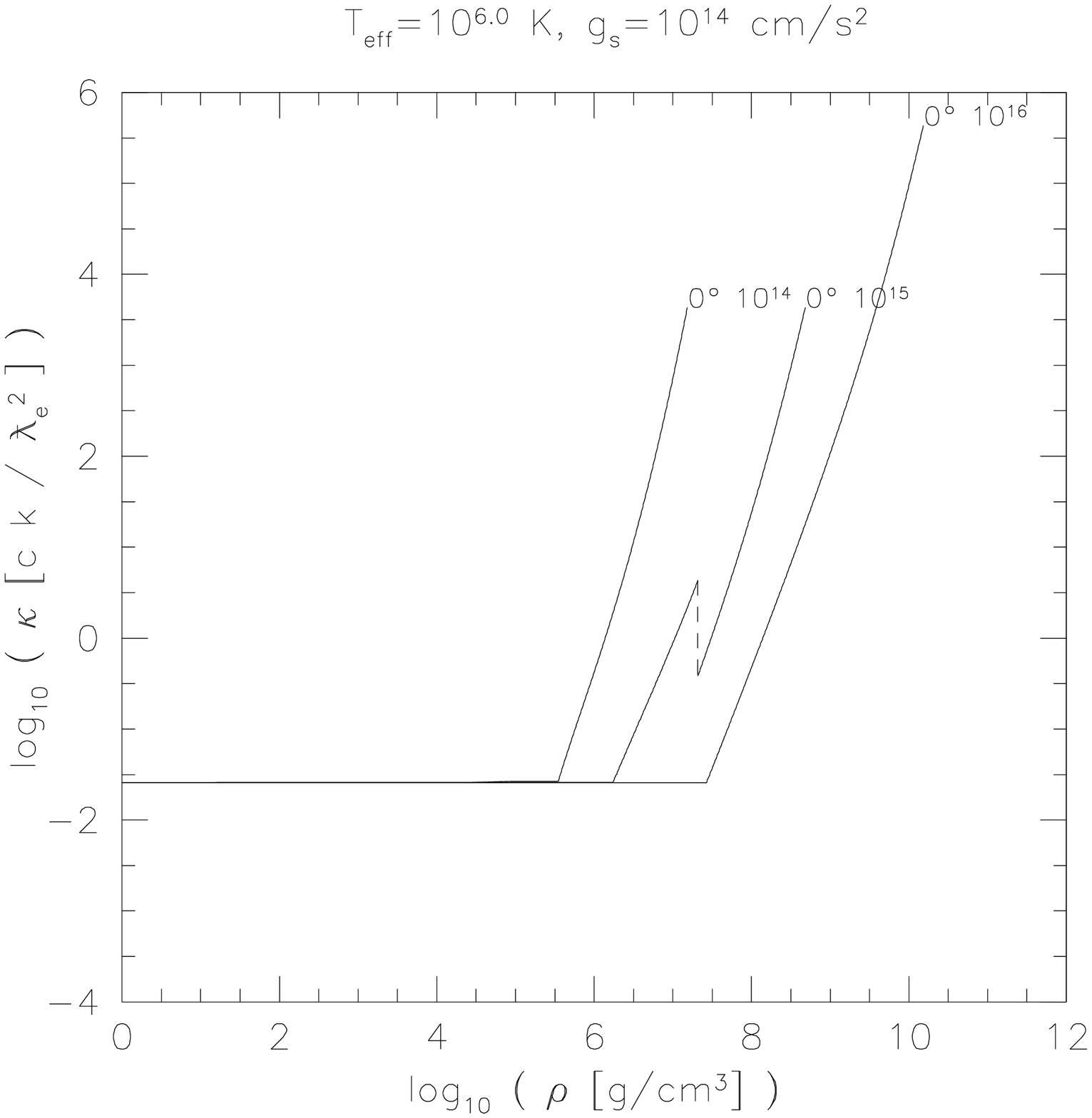}
\caption{Thermal structure of a strongly magnetized neutron star
envelope for a radial field.  The left panel traces the
temperature-density relation with $B=10^{14}, 10^{15}, 10^{16}$~G and
effective surface temperature of $10^6$~K.  The right panel traces the
conductivity through the envelope.  The constant conductivity solution
appropriate for a purely power-law conductivity law works well through
the nondegenerate regime.  In the left panel, liquid phase exists
above the dashed curve, and solid phase exists below.}
\label{fig:rhot6}
\end{figure}}
In \figref{rhot6} we present results for the degenerate and
non-degenerate regime for several magnetic fields with an effective
surface temperature of $10^6$ K.  In the nondegenerate regime the
temperature solution follows the power-law given in
\eqref{Trhond} and the {\it conductivity} is nearly constant.  In the
degenerate regime, the conductivity increases dramatically and the
{\it temperature} remains nearly constant.  For the solution with
$B=10^{15}$ G, the discontinuity in the conductivity at the phase
transition is apparent.  The results for $10^{14}$~G qualitatively
agree with the results of \jcite{Hern85} for this field strength.
Quantitatively however, we find that the conductivity in the
nondegenerate regime is thirty percent lower than the earlier result
of $10^{14}$~erg/(K cm s) and is given by \eqref{kappasolnnd}.  As the
magnetic field strength increases, we find that the core temperature
(or here the temperature at which the first Landau level is filled)
decreases.  This effect results from the increased conductivity in the
nondegenerate regime where $\kappa$ is approximately proportional to
$\beta^2$ and the degenerate regime where the quantization of the
electron phase space increases the conductivity above the zero-field
values.

We take advantage of the simplicity of this analytic technique when
calculating the thermal structure for hotter and cooler surface
temperatures. We do not need to reintegrate the structure equations
themselves.  All that is required is to recalculate the boundary
conditions at the non-degenerate-degenerate interface and the
liquid-solid phase transition. Again we find qualitative agreement
with \jcite{Hern85}.  In the nondegenerate regime, the increased or
decreased flux mimics the effect of changing the field strength
depicted in \figref{rhot6}.

We compare the various results by determining the temperature at the
following densities: $\rho = 1.5 \times 10^7$, $4.7 \times 10^8$ and $1.5
\times 10^{10}$~g/cm$^3$.  These are the densities at which the lowest
Landau level fills for field strengths of $10^{14}$, $10^{15}$ and
$10^{16}$~G.  Moreover, since the matter is nearly isothermal at higher
densities, these temperatures are close to the core temperature, at
least for parallel transport. By fitting the results of the
calculations, we find that at the lowest density:
\be
T(\rho=1.5 \times 10^7 \rmmat{g/cm}^3) \propto \beta^{-0.19} 
\left ( \frac{F}{g_s} \right )^{0.35}
\label{eq:tclof}
\ee
and at both the higher densities
\be
T(\rho=4.7 \times 10^8 \rmmat{g/cm}^3,1.5 \times 10^{10} \rmmat{g/cm}^3) 
\propto \beta^{-0.16} 
\left ( \frac{F}{g_s} \right )^{0.43}.
\label{eq:tchif}
\ee
\figcomment{\begin{figure}
\plottwo{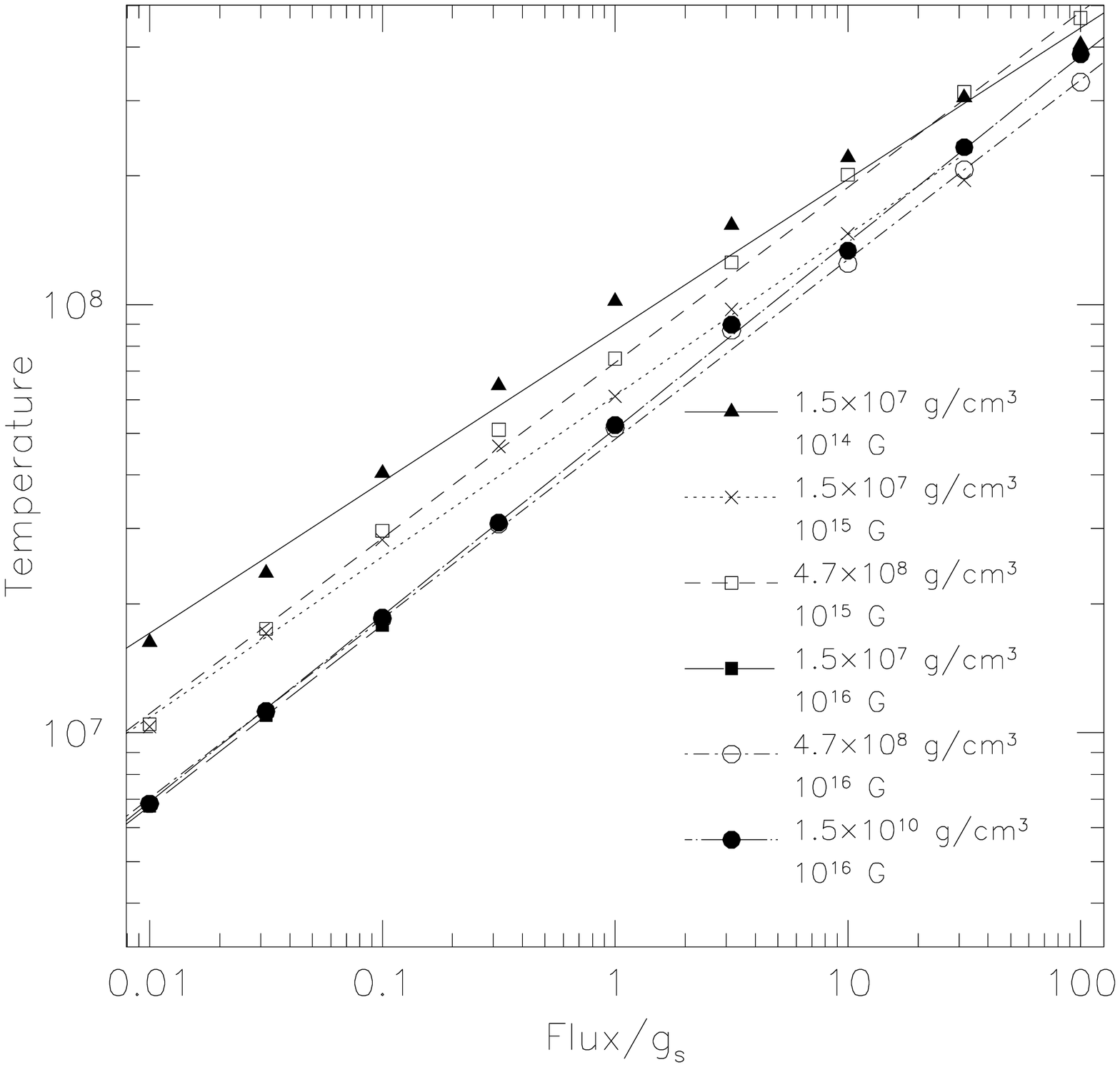}{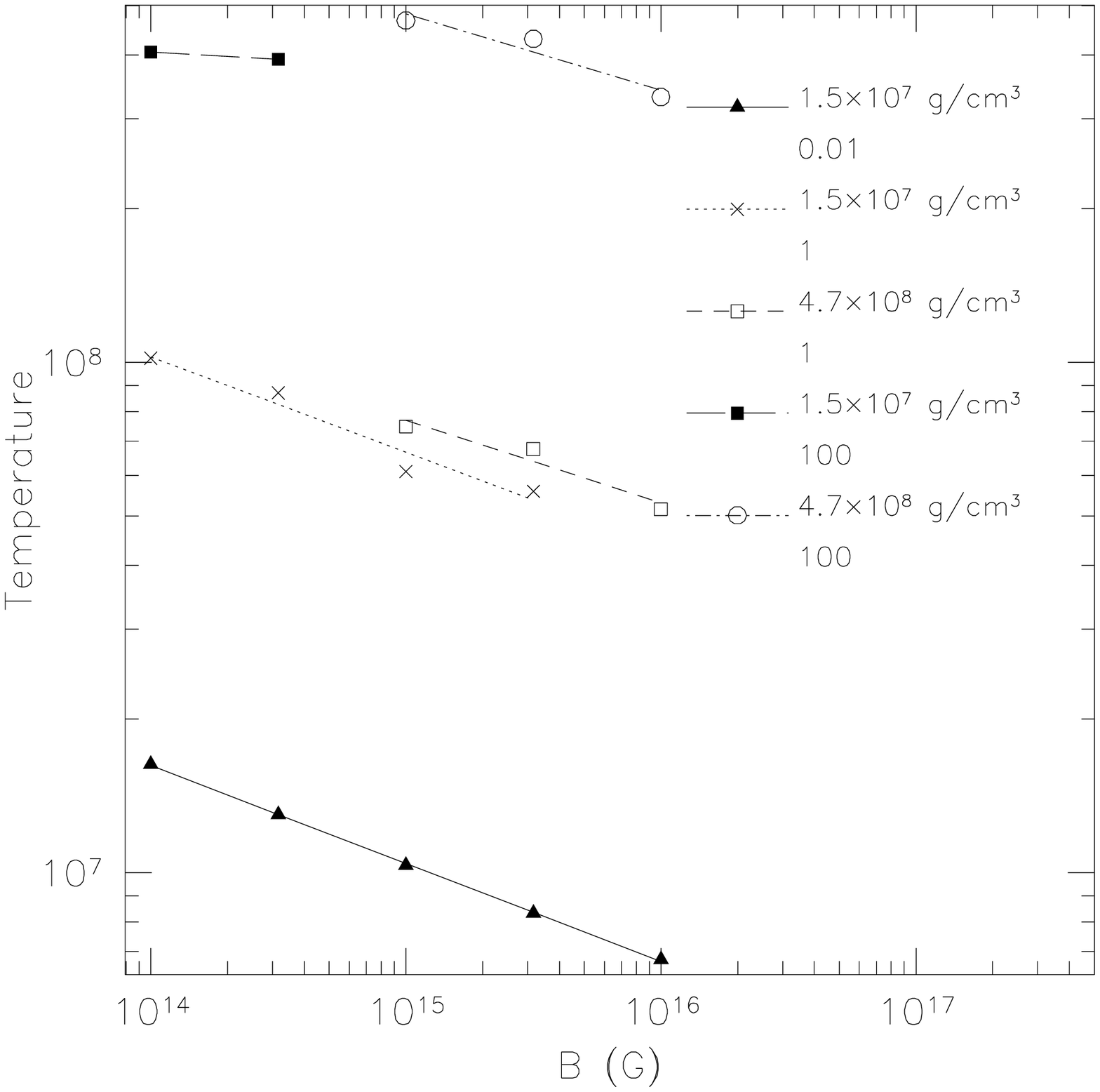}
\caption{The left panel depicts the temperature-flux relation for
several field strengths and densities.  $F/g_s$ is given in units of 
$\sigma (10^6 K)^4/10^{14}$~cm/s$^2$.
The right panel depicts the
temperature-magnetic-field relation.  The symbols show the calculated
data points and the lines are the best-fit power law functions to the data.}
\label{fig:Tcfits}
\end{figure}}
\figref{Tcfits} compares the numerical results with the best-fit
power-law relations.  

\paragraph{Perpendicular Transport}

\figcomment{\begin{figure}
\plottwo{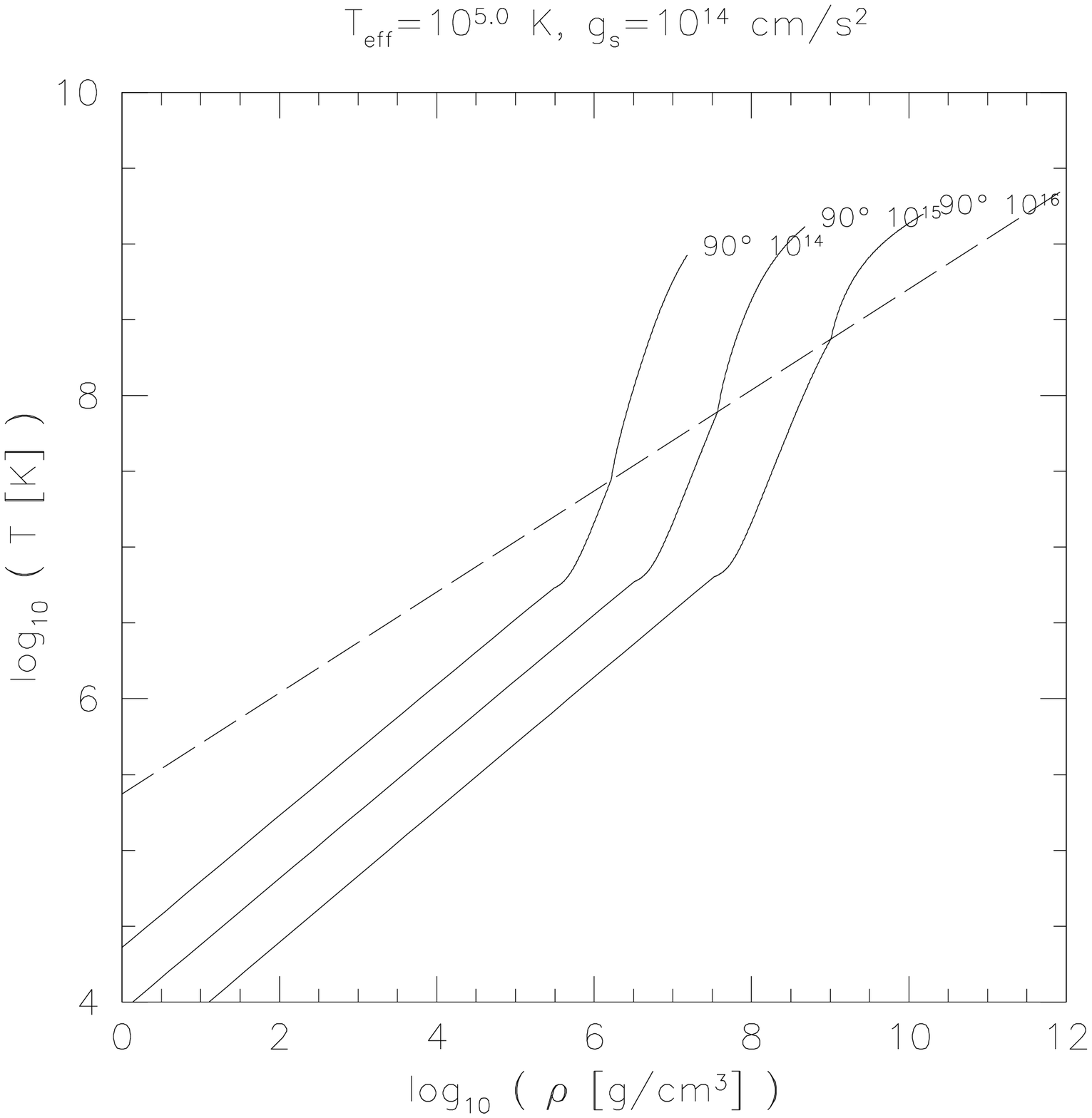}{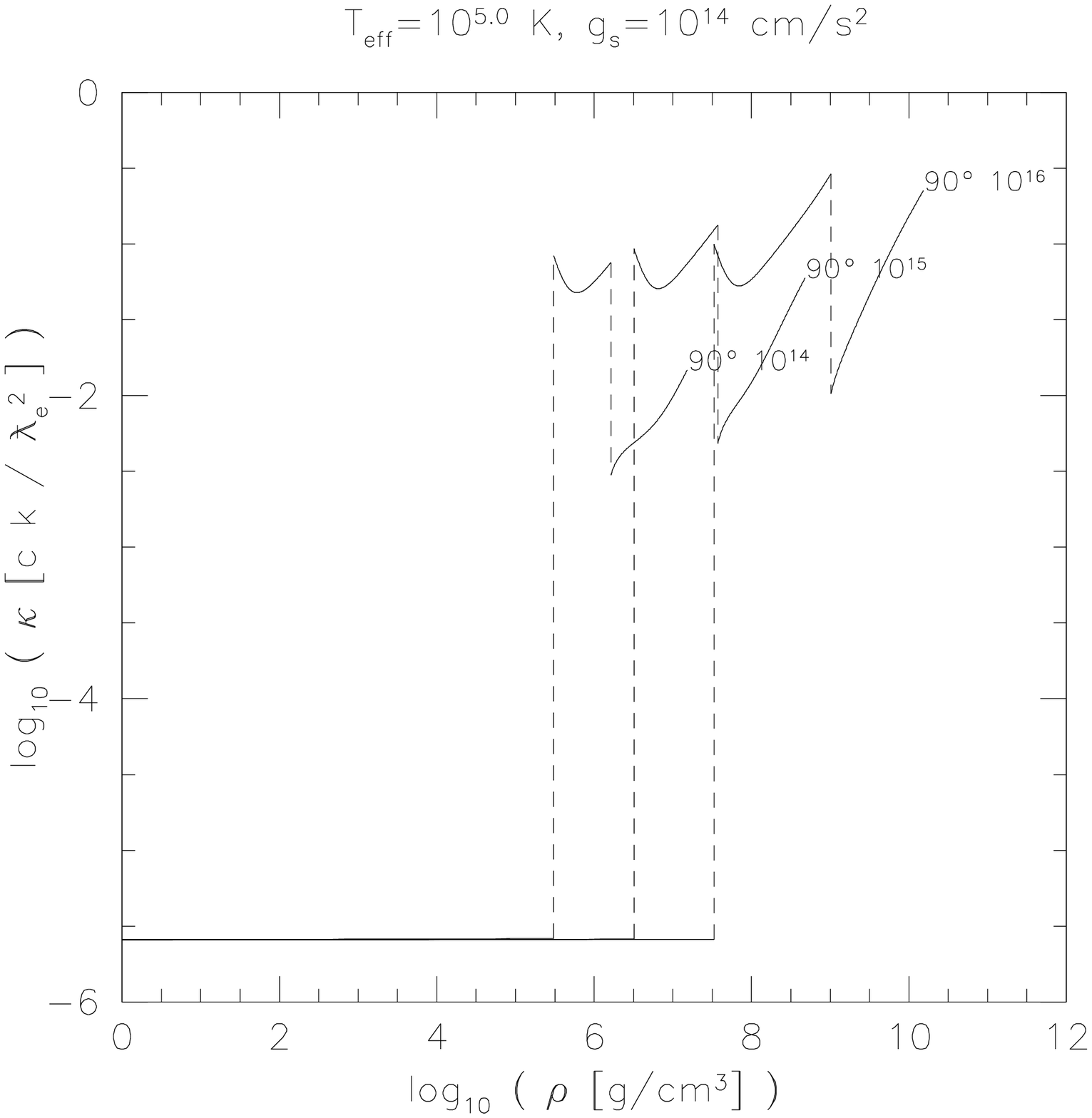}
\plottwo{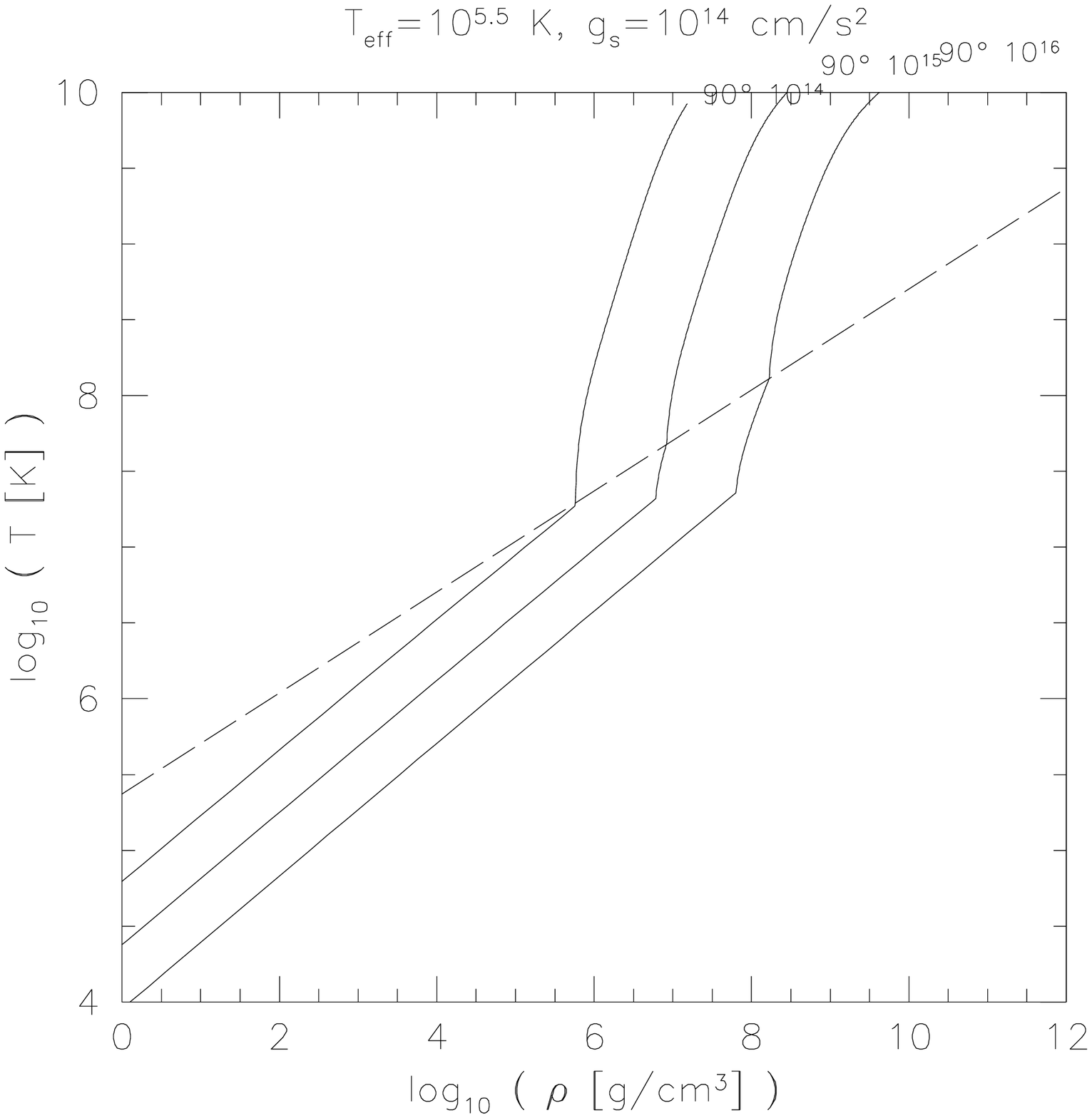}{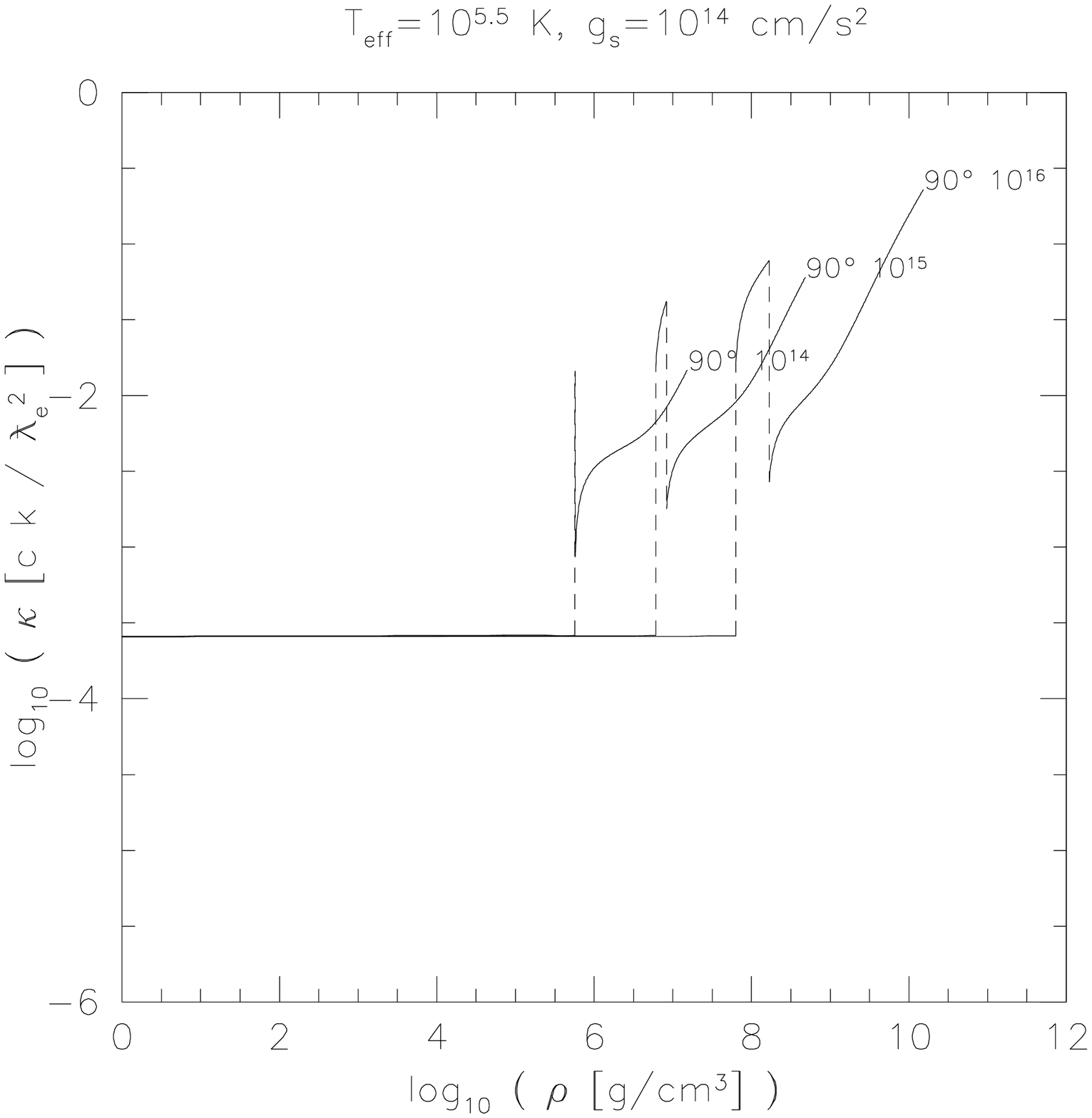}
\caption[0]{Same as \figref{rhot6}
 for the perpendicular case.}
\label{fig:rhot5_90}
\end{figure}}

Modeling the transition between photon and electron heat transport is
qualitatively different for transport perpendicular to the field lines.
In the parallel case, the conductivity from electrons typically
increases rapidly with density, and the transition from photon to
electron-dominated heat transport is abrupt.  For perpendicular
transport, the function $Q$ decreases with energy, and therefore the
conductivity decreases with density for fixed temperature.  In this case
the transition is more subtle.  Fortunately, the solution does not
depend strongly on how this transition is treated, so we
choose to employ $\zeta-1 > \tau$ to delineate the region where electron
conduction dominates.  The conductivity is not continuous
across this transition as is apparent in \figref{rhot5_90}.

We varied the definition of the non-degenerate-degenerate interface
and found that it had little effect on the 
$T_\rmscr{max}-T_\rmscr{eff}$ relation.
\figref{ndshift} shows how the solution changes if we move the
interface to a factor of ten higher or lower temperature (\ie
$\zeta-1>10 \tau$ and $\zeta-1>\tau/10$).  Although near the interface
the solutions differ dramatically, at higher densities the choice has
little effect.  The boundary condition at the transition is
unimportant for perpendicular transport, because the temperature rises 
quickly with density, and the solution quickly ``forgets'' the
boundary conditions, in a manner analogous to the convergence of the
radiative zero solution to the true solution in stellar atmospheres
(\eg \cite{Schw58}).  This is in contrast to the case where
$\psi\neq \pi/2$ where the material quickly becomes isothermal
in the degenerate regime.
\figcomment{\begin{figure}
\plotone{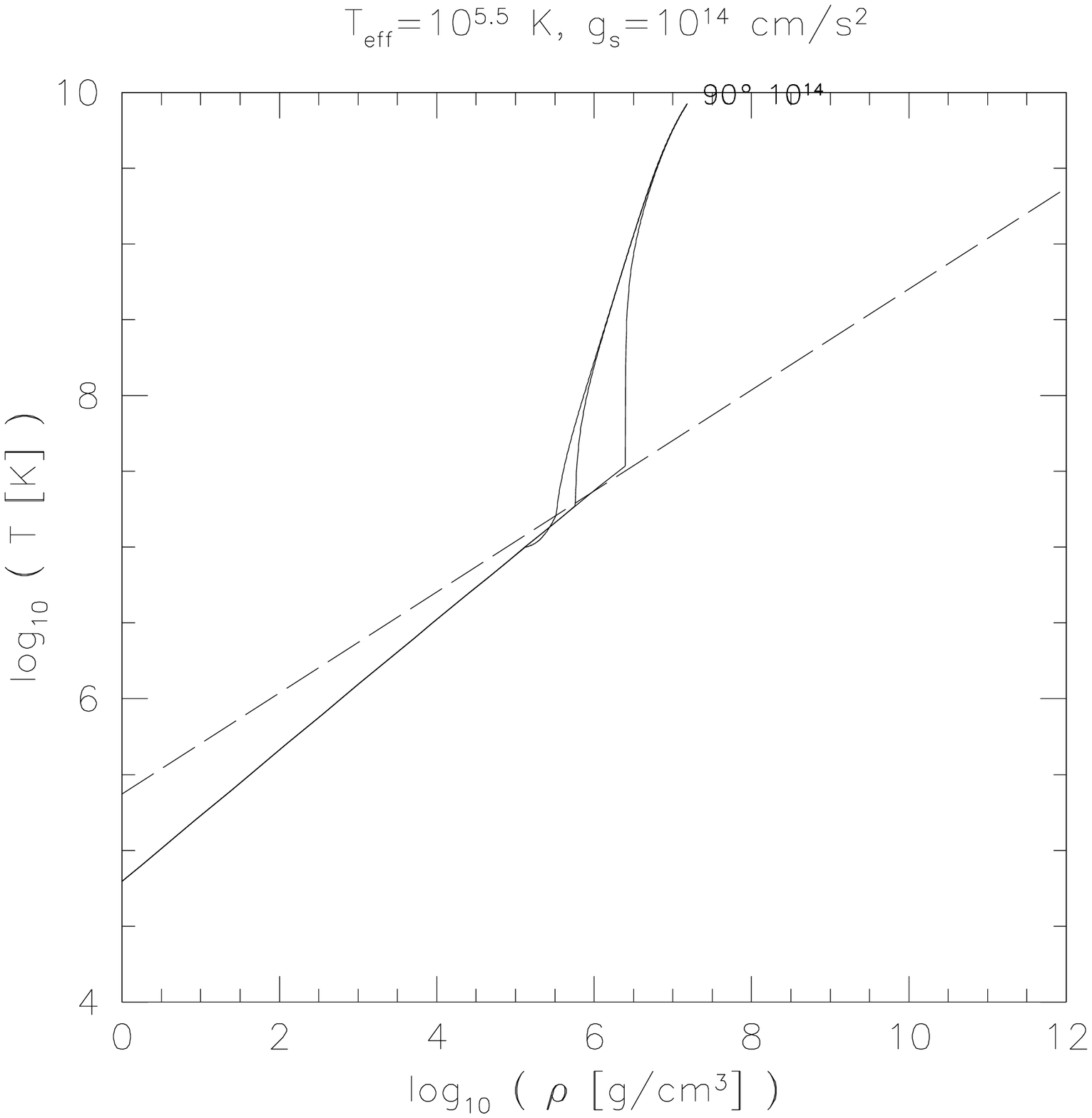}
\caption{The dependence of the envelope solution 
for transport perpendicular to the magnetic field upon the definition
of the non-degenerate-degenerate interface.  We have calculated the
location of the interface
for $(\zeta-1)/\tau = 0.1, 1, 10$ from left to right.}
\label{fig:ndshift}
\end{figure}}

We find that for a given effective temperature the core
temperature is much higher where the heat must travel perpendicularly
to the field lines.  Furthermore, we find that for stronger field
strengths the effect is more pronounced.

\paragraph{Effective Temperature Distributions} To find the effective
temperature as a function of angle with respect to the magnetic field we
vary the effective temperature as a function of angle until the
temperature where the first Landau level fills is constant for the
various angles.  Unfortunately, where the magnetic field is neither
radial or tangential we have solutions only in the non-degenerate and
liquid degenerate regimes.  Therefore, for acute angles we must select 
fluxes such that the degenerate solution is entirely in the liquid
regime.  For the more strongly magnetized envelopes we can follow the
solution to higher densities; consequently, we must use larger
effective temperatures for the stronger magnetic fields.  \tabref{twod}
summarizes the parameters for the calculations.
\figcomment{\bt
\caption{Results of the Two Dimensional Calculations}
\label{tab:twod}
\medskip
\begin{tabular}{ccc}
\hline
\multicolumn{1}{c}{Field Strength [G]} &
\multicolumn{1}{c}{$T_\rmscr{eff}(\psi=0)$ [K]} &
\multicolumn{1}{c}{$T_\rmscr{max} (\zeta=\sqrt{2\beta+1})$] [K]}
\\ \hline
$10^{14}$ & $1.07 \times 10^6$ & $1.12 \times 10^8$ \\
$10^{15}$ & $3.16 \times 10^6$ & $4.79 \times 10^8$ \\
$10^{16}$ & $5.06 \times 10^6$ & $8.13 \times 10^8$ \\
\hline
\end{tabular}
\et}
\figref{rhottang} depicts the results for $B=10^{14}$ and $10^{16}$~G.
For all but the perpendicular case, the envelope has become nearly
isothermal by the density where the first Landau level fills.

\figcomment{\begin{figure}
\plottwo{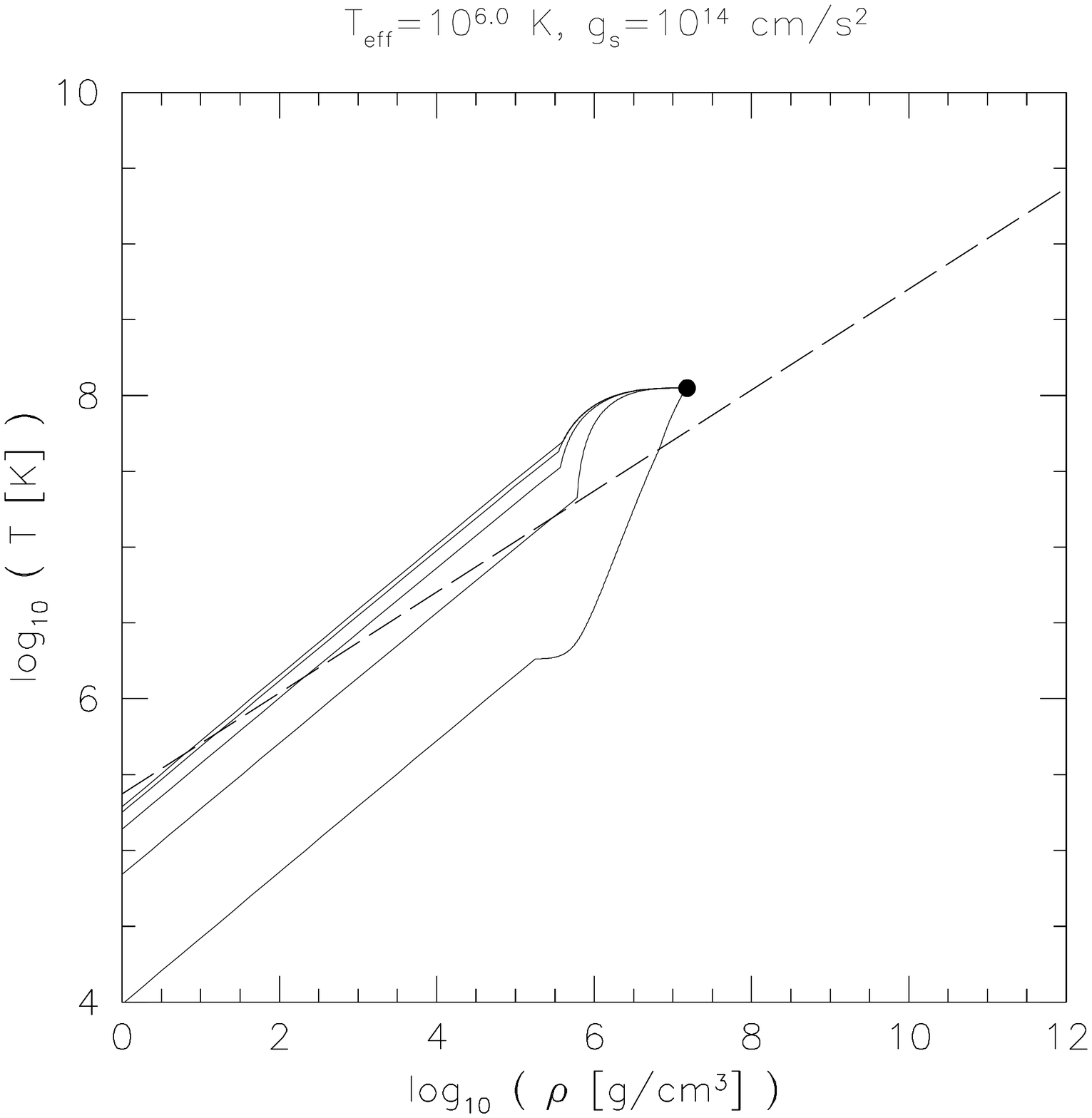}{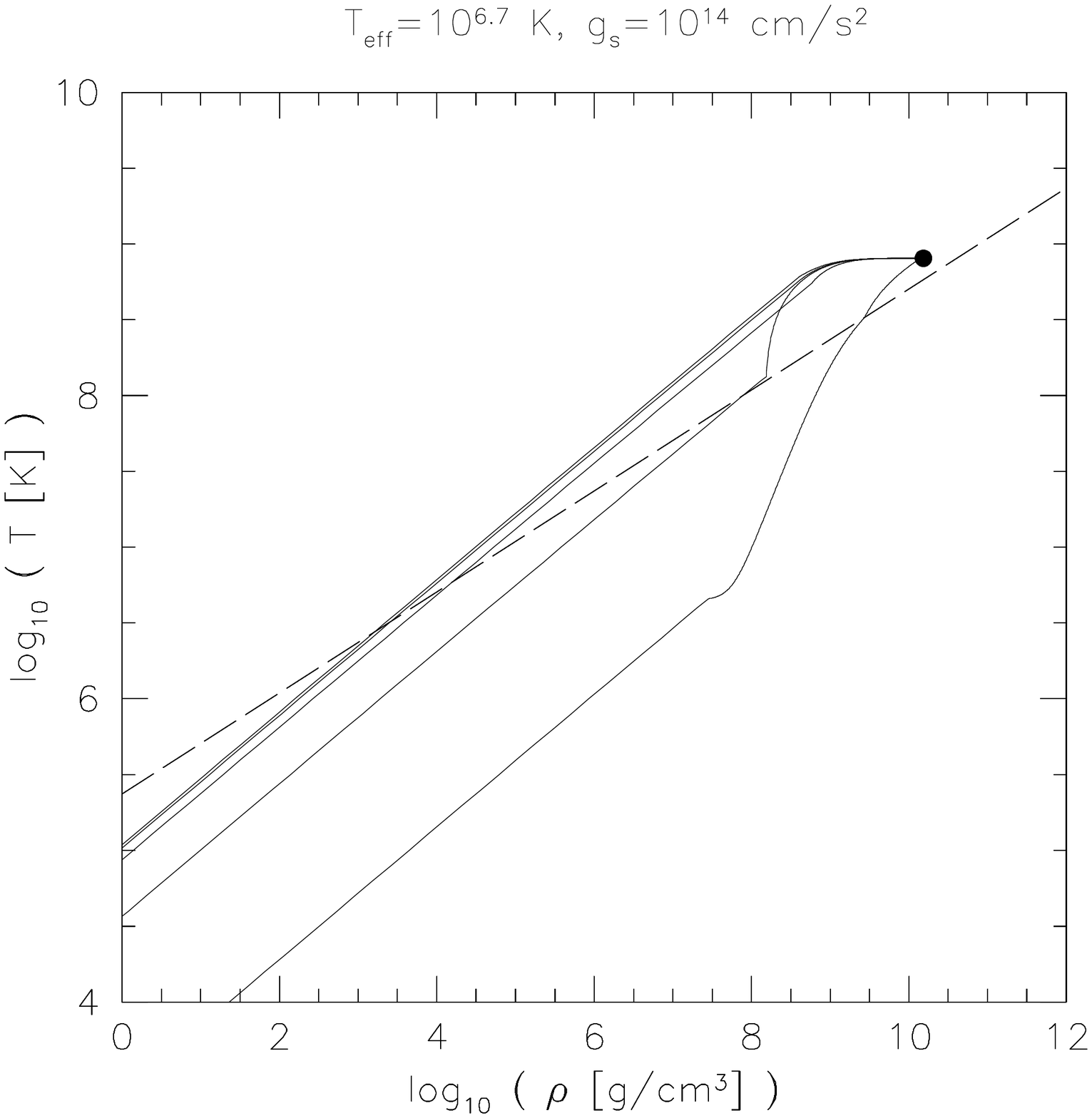}
\caption[gg]{The left panel shows the temperature structure of the envelope
as a function of density for $B=10^{14}$~G and $T_\rmscr{eff}=1.07 \times
10^6$ K.  From top to bottom, the results are for $\psi=0^\circ,
30^\circ, 60^\circ, 85^\circ, 90^\circ$ where $\psi$ is the angle between
the magnetic field and the radial direction.  The right panel depicts 
the temperature structure for $B=10^{16}$ G and $T_\rmscr{eff}=5.06
\times 10^6$ K.  The solutions are constrained to have the same
temperature at the density where the first Landau level fills (denoted
by the bold circle).  In both panels, liquid phase exists above the
dashed curve, and solid phase exists below.}
\label{fig:rhottang}
\end{figure}}

\figcomment{\begin{figure}
\plotone{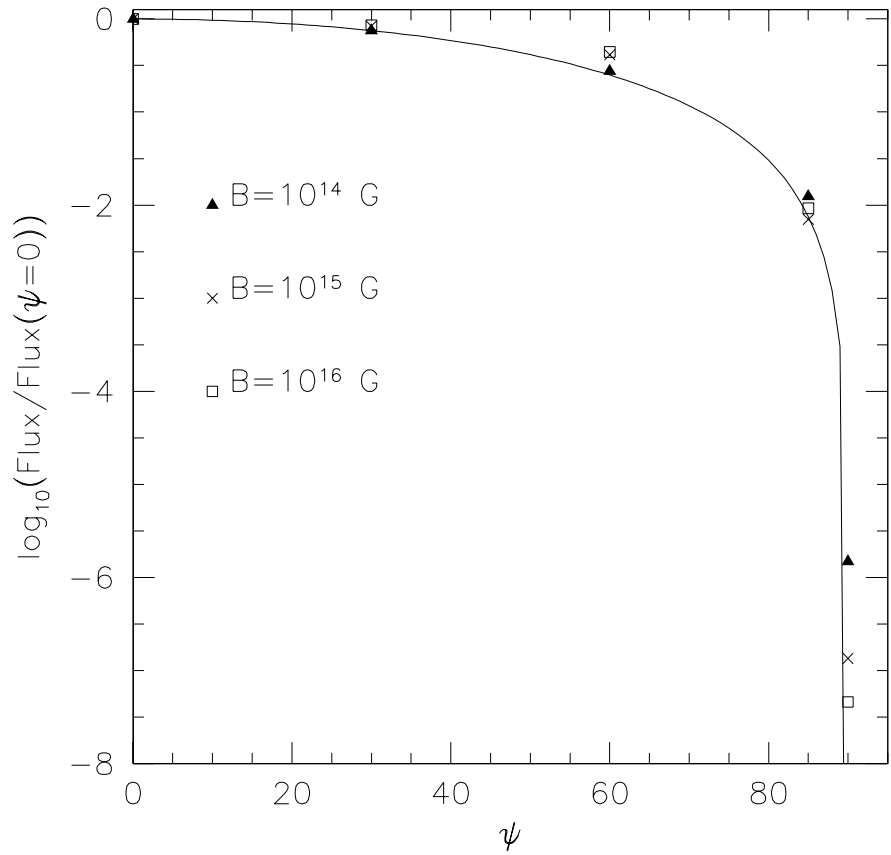}
\caption{Flux as a function of angle for $B=10^{14}, 10^{15}$ and
$10^{16}$~G.  The solid curve is $\cos^2 \psi$.}
\label{fig:fluxang}
\end{figure}}

\figref{fluxang} shows the flux as a function of angle for all of the
two-dimensional calculations.  The agreement between the flux
distribution and a simple $\cos^2 \psi$ law is striking.
\jcite{Gree83} have argued that if the conductivity is constant through
the envelope, the flux will follow a distribution of the form $A
\cos^2 \psi + B \sin^2 \psi$.  Although in the nondegenerate regime
the conductivity along a $T(\rho)$ solution is nearly constant, in the
degenerate regime it varies by several orders of magnitude.
Furthermore, the nondegenerate layers do not throttle the heat flux;
if they did, one would expect little variability, as the conductivity
parallel and perpendicular to the field are nearly equal in the strong
field limit. 

We look to the degenerate structure equation for the liquid state to
explain the remarkable agreement with a $\cos^2 \psi$ distribution.  
>From examination of \eqref{dtaudzetaliquid}, we see that if the
conductivity transverse to the field is neglected, we can make the
replacement 
\be
F \rightarrow F \cos^2 \psi
\ee 
and recover the thermal structure equation for $\psi=0$.  We determine where this
approximation is valid by comparing the
transverse and parallel components of the conductivity tensor
\be
\frac{\kappa_{yy,ei}}{\kappa_{zz,ei}} = \frac{Z^2 \alpha^4}{4 \pi^2 \beta^3}
 (\zeta^2-1) \frac{Q_{ei}(\zeta;\beta)}{\phi_{ei}(\zeta;\beta)} = 8.22
\times 10^{-11} \frac{Z_{26}^4}{A_{56}^2 B_{14}^5} \rho_6^2 
\frac{Q_{ei}(\zeta;\beta)}{\phi_{ei}(\zeta;\beta)}.
\ee
At first glance, it appears that the transverse conductivity is
negligible throughout the degenerate regime.  However, the functions
$Q_{ei}$ and $\phi_{ei}$ complicate the discussion.  Specifically,
$\phi_{ei} \rarrow 0$ and $Q_{ei} \rarrow \infty$ as $\zeta \rarrow 1$;
therefore, transverse conduction is likely to be important in the
nonrelativistic portion of the degenerate envelope.  \figref{rhottang}
shows that this is indeed the case.  For $B=10^{14}$~G, the solutions
for $\psi<\pi/2$ are nearly identical for $\rho > 10^{6.5}$~g/cm$^3$ or
$\zeta > 1.1$.  As $\zeta$ approaches unity, the ratio of the
conductivities increases without bound, the transverse conductivity may
no longer be neglected, and the runs of temperature with density begin to
diverge.

Empirically, we find that in the region of the envelope which most
effectively throttles the flux, the transverse conductivity may be
neglected for $\psi<\pi/2$ without introducing significant error.

\subsection{Observed Flux Distribution}

We follow the technique outlined by \jcite{Page95} to calculate the
observed fluxes.  However, unlike \jcite{Page95} we evaluate the double 
integrals over the visible portion of the neutron star surface directly.  
We use the $\cos^2 \psi$ rule to calculate the photon distribution function 
at the surface, and so we do not define a grid of precalculated distribution 
functions as \jcite{Page95} does.

As a first approximation, we focus on the variation of the observed
bolometric flux with the angle $\varphi$ of the line of sight with the
magnetic dipole axis.  This angle is a function of the phase angle
($\gamma$), the inclination of the dipole to the rotation axis
($\alpha$) and the line of sight to the rotation axis ($\zeta$)
\be
\cos \varphi = \cos \zeta \cos \alpha + \sin \zeta \sin \alpha \cos \gamma
\ee
(\cite{Gree83}).
For simplicity, in the discussion that follows we will take
$\alpha=\zeta=\pi/2$; therefore $\varphi=\gamma$ and we refer to
$\varphi=0,\pi$ as on-phase and $\varphi=\pi/2,3\pi/2$ as off-phase.

We repeat the calculation for several values of the stellar radius (with
fixed mass) to determine the effects of general relativity on the light
curves: gravitational redshift and the deflection of null geodesics
(self-lensing, or more concisely ``lensing'').  

To quantify the effect of gravitational lensing on the
light curves of magnetized neutron stars, we calculate the mean value
of the bolometric flux emitted by the surface over the visible region of
the star.  We assume that the flux at a given location on the
surface is proportional to $\cos^2\psi$ where $\psi$ is the angle
between the radial direction and the magnetic field.

For clarity, we treat the gravitational redshift separately.
\figref{lens1} depicts the ratio of the mean value of the flux over 
the visible portion of the star to the flux that would be emitted if
the magnetic field were normal to the surface throughout (\ie an
isotropic temperature distribution).
\figcomment{\begin{figure}
\plotone{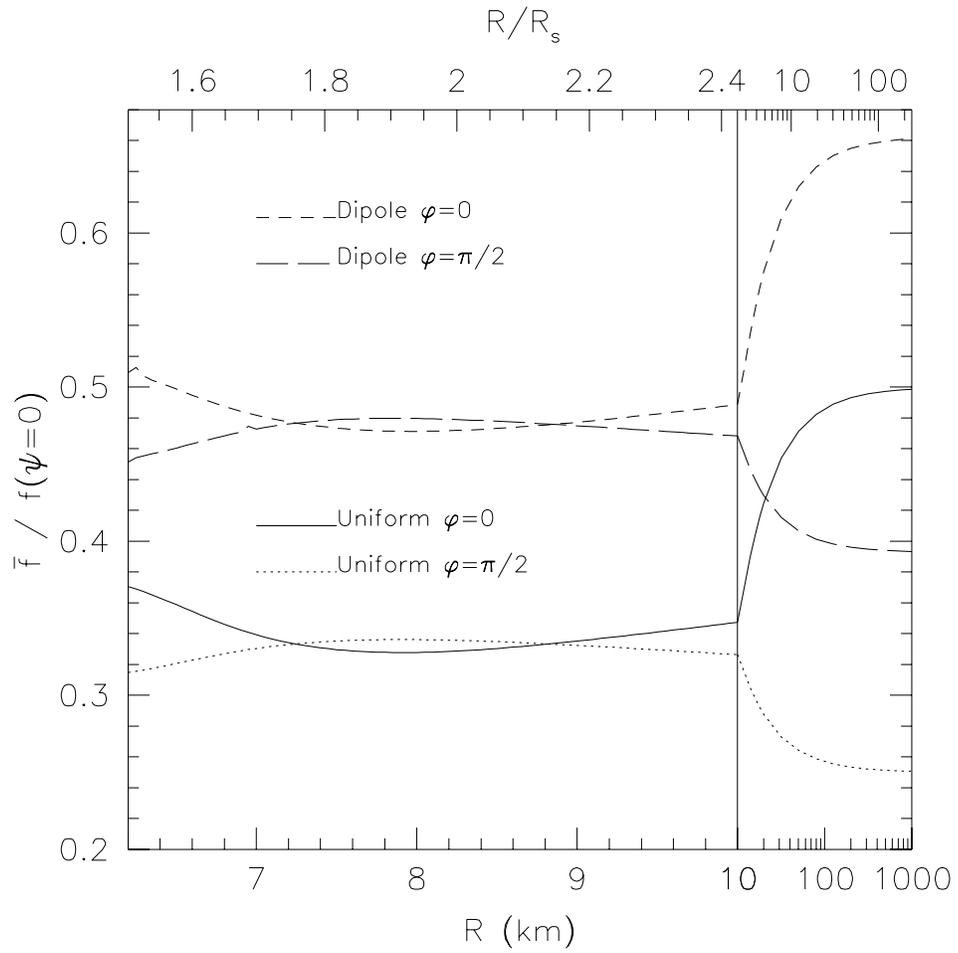}
\caption{The average of the bolometric flux over the visible 
portion of the neutron star for $\varphi=0,\pi/2$.  
The upper pair is for a dipole, and the lower
is a uniform field configuration.}
\label{fig:lens1}
\end{figure}}

In the limit of infinite radius, \ie if lensing is unimportant, we find
that for a uniform field
\be
{\bar  f} (\varphi=0) = \frac{1}{2} f (\psi=0) \rmmat{~and~}
{\bar f} (\varphi=\pi/2) = \frac{1}{4} f(\psi=0).
\ee
For a dipole field, the calculation is slightly more complicated.
First, we used \eqref{cos2psi} to determine the angle of the field
with respect to the radial direction.  Secondly from
equations~\ref{eq:tclof} and~\ref{eq:tchif}, we find that the
emergent flux is a function of the field strength.  For a dipole
configuration, the magnitude of
the field varies as
\be
\beta \propto \sqrt{3\cos^2\theta+1}
\ee
along the surface of the star.  Since we are most interested in fixing
the internal temperature at high densities we assume that the flux is
proportional to $B^{0.4}$ from \eqref{tchif}
which reduces the flux for $\theta\sim\pi/2$
further beyond the $\cos^2\psi$ rule.  We obtain
\be
{\bar  f} (\varphi=0) = 0.663 f (\psi=0) \rmmat{~and~}
{\bar f} (\varphi=\pi/2) = 0.393 f (\psi=0). \\
\ee
If we did not include the effect that the flux is a function of field
strength as well as orientation we would have obtained 0.717 and 0.444
for the above values.

We find that the mean flux is greater for the dipole configuration
than for a uniform field for all viewing angles if $R/R_s<5$, and that
the variation in the light curve is generally smaller.  We have taken
$M_\rmscr{NS}=1.4 \rmmat{M}_\odot$, yielding a Schwarzschild radius,
$R_s$, of 4.125 km.  The theoretical predictions for the radius of a
$1.4 \rmmat{M}_\odot$ neutron star range from 6.5 km to 14 km
(\cite{Thor94}; \cite{Pand71}; \cite{Wiri88}), depending on the details
of the equation of star at supernuclear densities.

As \jcite{Page95} found, lensing dramatically reduces the variation of
the observed flux with phase by making more than half the surface
visible at any time.  Interestingly, for the range of radii 7.248 --
8.853 km, we find that the flux is greater when the magnetic poles are
located perpendicular to the line of sight.  For this range of radii
over 90\% of the surface is visible.  Emission from
both of the hotspots reaches the observer leading to a larger flux.
For radii less than 7.248 km the entire surface is visible and again the
peaks are on phase.  \jcite{Page95} found a similar effect for the same
range of radii.

The emitted spectra from the visible portion of the neutron-star surface
is the sum of blackbody spectra of various temperatures.  To determine
the emitted spectra, we calculate the distribution of blackbody
temperatures on the surface; \ie we estimate the distribution function
$\d {\bar f}/\d T_\rmscr{eff}$.  With this distribution function, it is
straightforward to calculate the emergent spectrum averaged over the
visible portion of the surface as
\be
{\bar f}_\omega = \int_0^\infty \dd{\bar f}{T_\rmscr{eff}} 
\frac{1}{\sigma T_\rmscr{eff}^4}
\frac{\hbar}{4 \pi^2 c^2} \frac{\omega^3}{\exp(\hbar \omega/k T)-1} 
d T_\rmscr{eff}.
\ee
The factor of $\sigma T_\rmscr{eff}^4$ converts the flux to an effective
area of emission ($\bar A$).  The calculation of $\d {\bar f}/\d
T_\rmscr{eff}$ is numerically more tractable than $\d {\bar A}/\d
T_\rmscr{eff}$ and allows us to account for the total energy emitted
more reliably. 

We calculate $\d {\bar f}/\d T_\rmscr{eff}$ in similar fashion to $\bar
f$.  To expedite the calculation, we note that given the $\cos^2\psi$ rule 
the neutron star surface has a limited range of effective
temperatures, specifically between 0 and $T_\rmscr{eff}(\psi=0)$.
Consequently we define
\def\Tm{{\tilde T}}
\def\dfdTm{{\dd{\bar f}{\Tm}}}
\def\dfodTm{\d {\bar f}/ \d {\Tm}}
\ba
\Tm &=& \frac{T_\rmscr{eff}}{T_\rmscr{eff}(\psi=0)} \\
\dfdTm &=& \dd{\bar f}{T_\rmscr{eff}} T_\rmscr{eff}(\psi=0). 
\ea
For observations on-phase ($\varphi=0$) and without lensing ($R\rarrow\infty$),
the flux-weighted temperature distribution can be calculated directly if the
envelope is uniformly magnetized.  It is given by
\be
\dfdTm = 4 \Tm^7 f(\psi=0). 
\label{eq:dfdTmu}
\ee
The result for the dipole cannot be written explicitly and is not
illustrative.

For a general geometry ($\varphi \neq 0$), we expand this function in
an orthonormal basis on the interval 0 to 1. 
Specifically, we assume that for $0\le \Tm\le 1$
\be
\dd{\bar f}{\Tm} = 
\sum_{l=0}^\infty A_l Q_l(\Tm)
\ee
and zero otherwise, where
\be
Q_l(\Tm) = \sqrt{2l+1} P_l(2\Tm-1).
\ee
The $P_l(x)$ are the Legendre polynomials.  From
the properties of these orthonormal functions, we have $A_0={\bar f}$,
and $A_l$ is calculated by inserting the weighting function $Q_l(\Tm)$
into the integrands in the calculation for $\bar f$.  We recall that
we have assumed, $\Tm=\cos^{1/2}\psi$

Using an orthonormal basis dramatically speeds the calculation of the
distribution.  Additionally, because the $P_l(x)$ are polynomials, it
is straightforward to calculate conventional statistics of the
distribution
\ba
\left < \Tm \right > &=& \frac{1}{\bar f} \int_0^1 \Tm 
\: \dd{\bar f}{\Tm} \, d \Tm 
 = \frac{\sqrt{3}}{6} \frac{A_1}{A_0} + \frac{1}{2} \\
\left < \Tm^2 \right > &=& \frac{\sqrt{5}}{30} \frac{A_2}{A_0} +
\frac{\sqrt{3}}{6} \frac{A_1}{A_0} + \frac{1}{3}.
\ea
Unfortunately, with this basis is impossible to insist that distribution
is everywhere non-negative, \ie that no temperatures contribute negative
flux.  However, if a sufficiently large number of $A_l$ are calculated,
the intervals where $\dfodTm<0$ can be made to be arbitrarily small and
to have an arbitrarily small contribution to the total flux.  We compare
the results of the expansion with Equation~\ref{eq:dfdTmu} and find that
the maximum relative error in the expansion coefficients between the two
techniques is approximately $9 \times 10^{-5}$. 

\figref{tdist} depicts the results of this calculation for four values
of the stellar radius with $\varphi=0,\pi/2$.  In the left panel, we
see in the absence of general relativistic effects that when the
neutron star is off-phase more flux is produced at lower blackbody 
temperatures than when the magnetic dipole is pointing toward the
observer.  For the smallest radius considered ($R=6.25$ km), the entire
surface of the neutron star is visible and a large portion of the front
hemisphere has a second image.  In this case, both the flux-weighted 
temperature distributions at $\varphi=0$ and $\pi/2$ are peaked at the maximum
effective temperature.  However, the distribution off-phase has a more well
populated tail extending toward lower temperatures than on-phase.
\figcomment{\begin{figure}
\plottwo{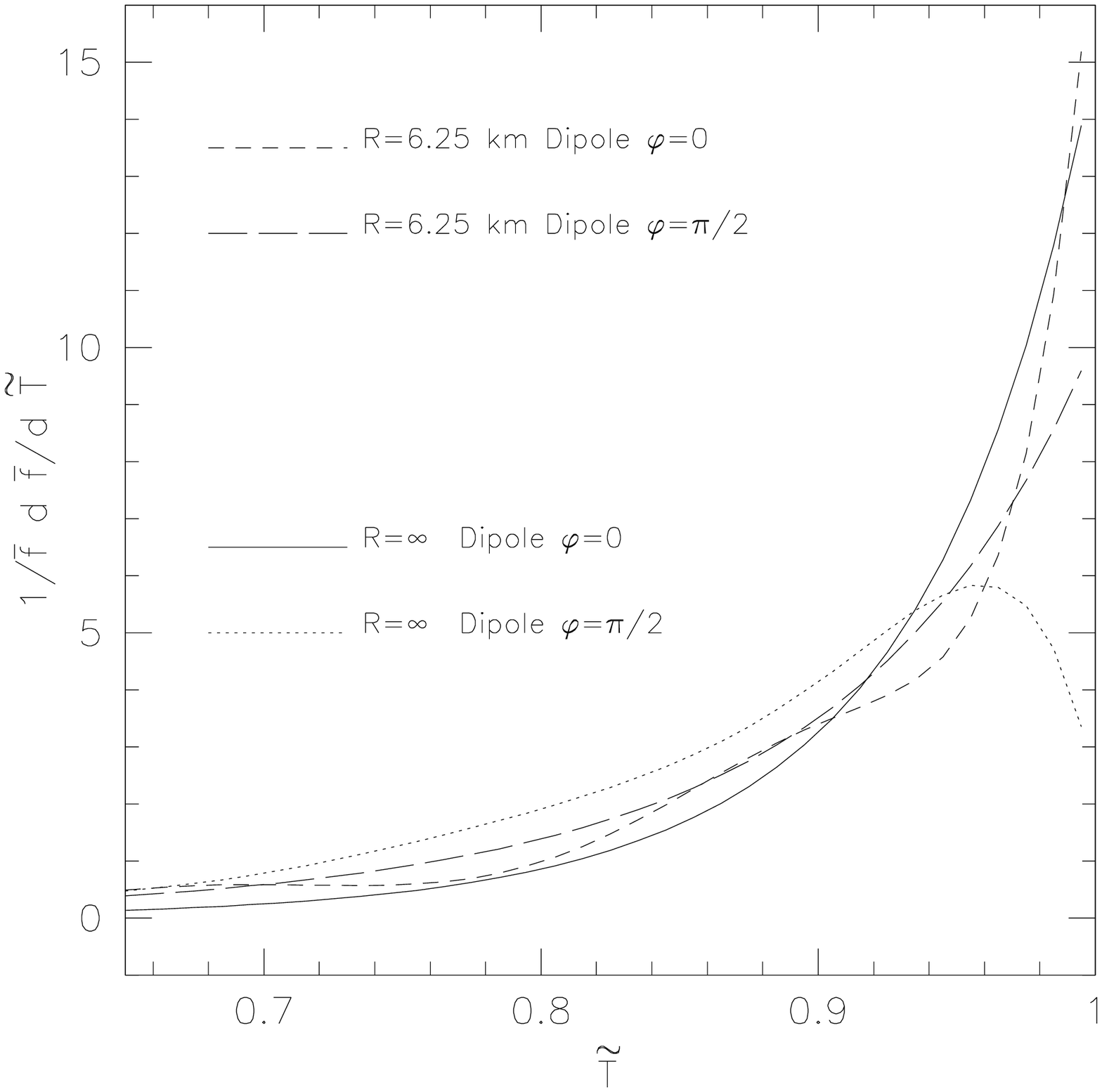}{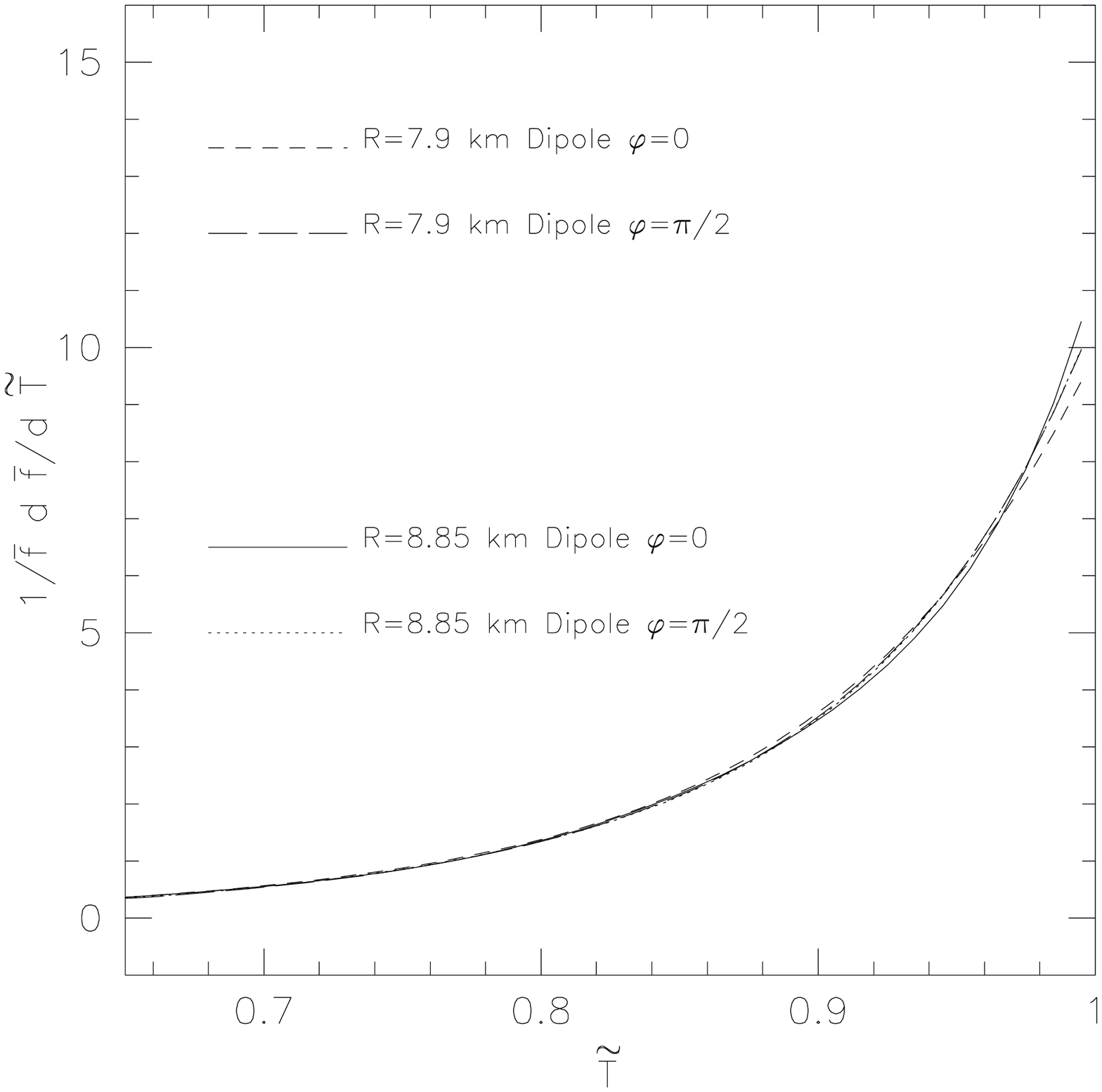}
\caption{The fractional distribution of observed flux as function of the
surface blackbody temperature.  The left panel depicts the
distribution for the minimal ($R/R_s=1.52$) and maximal radii
($R=\infty$) considered.  The right panel depicts the distribution at a
radii where ${\bar f}(\varphi=0)={\bar f}(\varphi=\pi/2)$ ($R=8.85$
km) and where the off-phase peaks are maximized ($R=7.9$ km).  All the
curves are normalized to have an integral of unity from $\Tm=0$ to 1}
\label{fig:tdist}
\end{figure}}

The right panel depicts a value of the stellar radius ($R=8.85$ km) 
where there is practically no variation of $\bar f$ with phase.
Additionally, we see that the flux-weighted temperature distributions
are nearly constant with phase.  Also depicted is the temperature
distributions for $R=7.9$ km, the radius where the off-phase peaks are
maximized.  Here again, the distributions do not change appreciably with
phase.  We conclude that for $7 \lsim R \lsim 9$ km, it would be
difficult to detect variation of the spectra with phase, if the neutron
star surface indeed radiates as a blackbody.

To calculate the spectra themselves, it is convenient to define the
following functions,
\def\Om{{\tilde \omega}}
\be
{\cal F}_l(\Om) = \frac{15}{\pi^4} \Om^3 
\int_0^1 Q_l(\Tm) \frac{1}{\Tm^4} \frac{1}{\exp(\Om/\Tm)-1} 
d \Tm,
\ee
where $\Om=\hbar \omega/ k T_\rmscr{eff}(\psi=0)$, 
so that
\be
{\bar f}_\omega (\omega)  =  \frac{\hbar}{k T_\rmscr{eff}(\psi=0)}
\sum_{l=0}^\infty A_l {\cal F}_l 
\left ( \frac{\hbar \omega}{k T_\rmscr{eff}(\psi=0)} \right ).
\ee
Since $A_l$ has units of flux, we obtain the correct units of flux-time
for ${\bar f}_\omega$.

To convert to the observed spectra, we must account for gravitational
redshift and interstellar absorption; we obtain (\cite{Page95})
\be
f^\rmscr{observed}_\omega (\omega) = {\bar f}_\omega (\expL \omega)
\expL \frac{R_\infty^2}{D^2} e^{-N_H \sigma(\omega)}
\ee
where
\be
R_\infty \equiv R e^{\Lambda_s}
\label{eq:ridef}
\ee
where $\expL$ is given in \eqref{expLdef}.  The final term accounts for
interstellar absorption and $D$ is the distance to the neutron star.
For $\sigma(E)$ we use the \jcite{Morr83} cross-sections.
\figcomment{\begin{figure}
\plotone{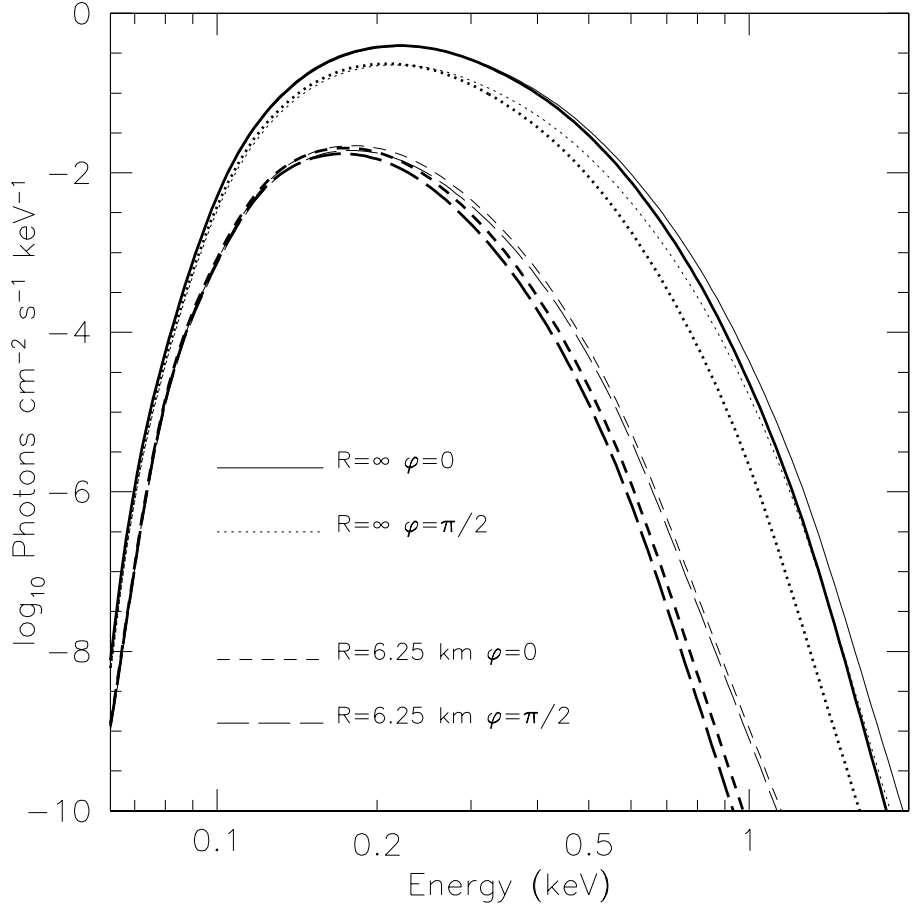}
\caption[]{The observed spectra from a neutron star
with $T_\rmscr{eff}(\psi=0)=7.5 \times 10^5$ K from a distance of 250 pc
with an intervening absorption column of $N_H=10^{20} \rmmat{cm}^{-2}$.
The light curves show the spectra, and the heavy curves show blackbody
spectra at the mean effective temperature of the neutron star.  For
the $R=\infty$ model, we have taken $R/R_s\rarrow \infty$ while $R=20$
km to give the surface area of the neutron star, while neglecting
general relativistic effects.}
\label{fig:compspec}
\end{figure}}

\figref{compspec} depicts the spectra in the ROSAT energy range for two
neutron stars.  Each of the spectra (light curves) is well fitted by a
blackbody (heavy curves) at the mean effective temperature with an
additional hard component.  The mean effective temperature
($T_\rmscr{mean}$) is defined by
\be
\int f^\rmscr{observed}_\omega(\omega) d \omega = \sigma T_\rmscr{mean}^4
\frac{R_\infty^2}{D^2}
\ee
for $N_H=0$, \ie it is the equivalent blackbody temperature that
accounts for all of the energy emitted from the neutron star surface.
The hard component is most significant when the star is observed at
right angles to the magnetic axis, and originates from the portions of
the hot polar caps that are visible even when the star is off-phase.

In Appendix \ref{sec:dataprod}, we present two {\tt XSPEC} models
which are available over the WWW.  With these models, one may simulate
observations from various x-ray instruments to estimate the observed
pulsed fractions for the models discussed in this section.  We give
an example in \figref{xspec}.

\subsection{Neutron Star Cooling}

We can use the results of the preceding sections to determine the
effects of the magnetic field on neutron star cooling rates.
Specifically, we take the ratio of the total flux from the surface with
and without a magnetic field for the same core temperature.  We have
used the results of \jcite{Hern84b}.  To determine the core temperature
for a given flux we combine their equations~(4.7) and~(4.8), switching
from the first relation to the second when the surface effective
temperature drops below $4.25 \times 10^5$ K.  The results do not depend
qualitatively on whether equation~(4.7)~or~(4.8) of \jcite{Hern84b} is
used. 

To compare our calculated temperature-flux relation with the results for
isotropic heat transport, we multiply the fluxes for the magnetized
envelopes by $0.4765$ to account for a dipole field configuration.
Additionally, we assume in our calculations that the envelope is
isothermal above the density ($\rho_\rmscr{max}$) 
at which the first Landau level fills and
use the temperature ($T_\rmscr{max}$) at this density to estimate the
flux in the unmagnetized case.  Only for the strongest field strength
considered ($B=10^{16}$ G) do our analytic calculations extend to the
core density assumed by \jcite{Hern84b} of $10^{10}$~g/cm$^3$; however,
we do not expect the results to be strongly sensitive to this cutoff
density as our solutions (for $\psi\neq \pi/2$) are nearly isothermal at
high densities. 
\figcomment{\begin{figure}
\plotone{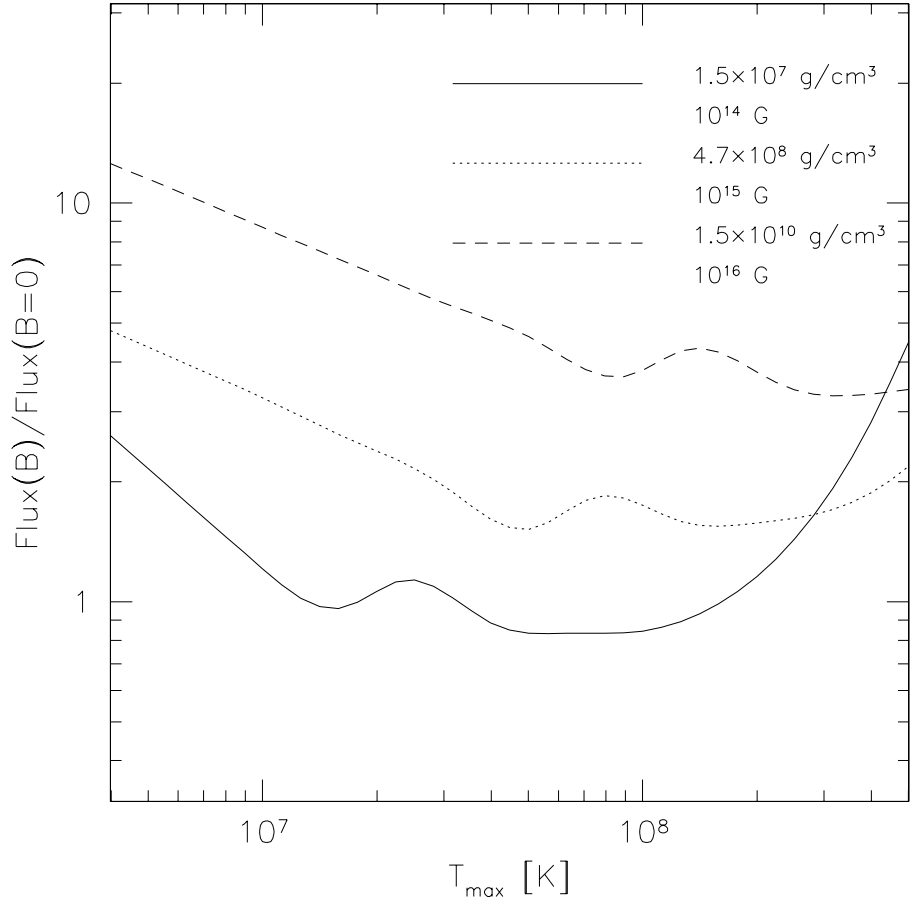}
\caption{The ratio of fluxes in magnetized envelopes to those in
unmagnetized ones.  We have assumed that the envelope is isothermal
above the densities given in the legend to estimate the unmagnetized 
fluxes.}
\label{fig:coolvstc}
\end{figure}}

\figref{coolvstc} depicts the results of this comparison.  We find that
for the weakest field strength considered ($B=10^{14}$~G), the magnetic
field has little effect on the total luminosity of the star.  However,
for cooler core temperatures and stronger magnetic fields, the
difference in the luminosities can be up to a factor of ten.  The flux
ratio is sensitive to the core temperature because
Equations~\ref{eq:tclof}~and~\ref{eq:tchif} have a slightly different
power-law index than the model assumed for the unmagnetized envelope
(0.392).  The inflection in each of the curves occurs when the material
near non-degenerate-degenerate interface melts as the core temperature
increases.  All the curves swing upward for high values of
$T_\rmscr{max}$ because for high fluxes (\ie high core temperatures) our
assumption that the temperature is constant for $\rho>\rho_\rmscr{max}$
no longer holds. 

Because strongly magnetized neutron stars emit significantly more
flux, we expect that the thermal history of magnetars should be
dramatically different from that of neutron stars with weaker magnetic
fields.  We discuss this issue further in \jcite{Heyl97magnetar}.

\section{Conclusions}

We have presented an analytic technique for calculating the thermal
structure of ultramagnetized neutron star envelopes.  We use the exact
thermal conductivities in an intense magnetic field of \jcite{Sila80}
and \jcite{Hern84a} in the non-degenerate and degenerate regimes,
respectively.  We make two simplifying approximations.  We assume that
the interface between degenerate and non-degenerate material is
abrupt.  \jcite{Hern85} numerically calculated the thermal structure
for $B=10^{14}$~G without this assumption.  Our agreement with this
earlier result shows that an abrupt interface is a good approximation.
Secondly, we use a standard simplification in the study of stellar
atmospheres which is to use the radiative zero solution to fix the
outer boundary condition (\cite{Schw58}).  Because the equation for
the thermal structure in the outermost layers is qualitatively similar
to the relation for an unmagnetized envelope, we conclude that as in
the unmagnetized case (\cite{Hern84b}), this is also an accurate
approximation.

A distinct approach is to treat the entire problem numerically,
which allows us to estimate directly the possible errors that our
simplifying assumptions introduce and alleviates the problem of how to
treat the regions where more than one Landau level is filled.
Numerical models for parallel and transverse conduction for
$B=10^{12}-10^{14}$~G and $\rho < 10^{10}$~g/cm$^{-3}$ are presented
in \jcite{Heyl98numens}.

We find that the relation between transmitted flux, core temperature
and field strength may be approximated by a power law and that the
effective temperature is proportional to $\cos^{1/2} \psi$ where
$\psi$ is the angle between the radial direction and the local
direction of the magnetic field.  Using the geometric result, we
calculate the observed spectra as a function of viewing angle
including the effects of general relativity for dipole and uniform
field configurations.  We extend the conclusions of previous work.  If
the surface is assumed to radiate as a blackbody and neutron stars
have radii within the currently accepted range, the anisotropic heat
transport induced by a dipole field configuration is insufficient to
produce the observed pulsed fractions even for ultramagnetized
envelopes.

\jcite{Pavl94} argue that in addition to the transmission of heat
through the envelope, the emission at the surface is also strongly
anisotropic.  Anisotropic emission can naturally produce large pulsed
fractions even when the temperature on the surface is uniform
(\cite{Zavl95,Shib95}).  Additionally, the composition of the
atmosphere may have a profound effect on the emergent radiation.  A
magnetized iron atmosphere produces substantial limb darkening for
$\psi \sim \pi/2$, and the decrement is strongest at high energies
(\cite{Raja97}).  It is straightforward to graft these atmospheres
onto the thermal envelopes calculated here to obtained the observed
time-dependent spectra for a variety of realistic neutron star models.
These effects along with anisotropic conduction may be sufficient to
account for the large observed pulsed fractions.

\section*{Acknowledgements}

We would like to thank the referee, Vadim Urpin, for valuable
suggestions.  This work was supported in part by a National Science
Foundation Graduate Research Fellowship, the NSF Presidential Faculty
Fellows program and Cal Space grant CS-12-97.

\bibliography{ns}
\bibliographystyle{jer}

\begin{appendix}
\section{{\tt XSPEC} Models}
\label{sec:dataprod}

Rather than present results for specific instruments and band passes, we
supply our results in machine-readable form.  We have calculated neutron
star spectra for several values of $R/R_s$, and $\varphi=0,\pi/2$.  We
assume a dipole field configuration and the $\cos^2\psi$ rule.  The
model is calculated for 
\be
T_{\rmscr{eff},\infty}(\psi=0) = T_\rmscr{eff}(\psi=0)
\sqrt{1-\frac{R_s}{R}} = 10^6 \rmmat{K}
\ee
where $R_s=2 G M/c^2$.  $T_{\rmscr{eff},\infty}(\psi=0)$ may be varied by
applying a redshift and renormalization to the spectra.  For ease of use
by the x-ray astronomy community, we have created {\tt XSPEC} table
models with the data.

Within {\tt XSPEC}, the models may be convolved with the response
matrix for various x-ray instruments and compared with observed
spectra.  By using the redshift ($z$) and normalization
($K$) parameters we may obtain models for different effective
temperatures and radii:
\ba
K &=& \left ( \frac{R_{\infty,\rmscr{km}}}{D_{10}} \right )^2
\left ( \frac{T_{\rmscr{eff},\infty}}{10^6 \rmmat{ K}} \right )^3  \\
z &=& \frac{10^6 \rmmat{ K}}{T_{\rmscr{eff},\infty}} - 1
\ea
where $R_{\infty,\rmscr{km}}$ is the source radius in km as observed
at infinity, \ie
\be
R_\infty = R e^{\Lambda_s} = R \left ( 1 - \frac{R_s}{R} \right )^{-1/2}
\ee
from \eqref{ridef} and $D_{10}$ is the distance to the neutron star in
units of 10 kpc.  This particular choice is consistent with the
definition of the {\tt bbodyrad} model in {\tt XSPEC}.

Each additive model contains the single interpolation parameter
$R_s/R$ which ranges from 0 to 0.6601.  As an illustration
\figref{xspec} depicts one of the models convolved with the ROSAT PSPC
response function for $R/R_s\rarrow\infty$, $R=20$ km and $D=250$ pc
with $N_H=10^{20}$ cm$^{-2}$.  Here, we have used the {\tt wabs} model
to calculate the interstellar absorption which assumes the
\jcite{Morr83} cross-sections.

The errorbars are calculated for an exposure of $10^4$ seconds.  For
these parameters, the variation of the thermal flux with phase is
apparent in the spectra.  However, as we saw in the previous sections
as $R/R_s$ decreases, the variation in the thermal flux weakens.

The table models are available at the following URLs: 
\\
\href{http://www.cco.caltech.edu/~jsheyl/analytic_ns/p0.fits}{{\tt{
http://www.cco.caltech.edu/$\sim$jsheyl/analytic\_ns/p0.fits}}}  for $\varphi=0$ \\
 \smallskip
\href{http://www.cco.caltech.edu/~jsheyl/analytic_ns/p90.fits}{{\tt{
http://www.cco.caltech.edu/$\sim$jsheyl/analytic\_ns/p90.fits}}} for
$\varphi=\pi/2$.
\\
The {\tt{XSPEC}} software itself is available at
\\ 
\href{ftp://legacy.gsfc.nasa.gov/software/xanadu/}
{{\tt{ftp://legacy.gsfc.nasa.gov/software/xanadu/}}}, 
\\ 
and an online manual is provided at 
\\ 
\href{http://www.merate.mi.astro.it/~xanadu/xspec/u_manual.html}{{\tt{
http://www.merate.mi.astro.it/$\sim$xanadu/xspec/u\_manual.html}}}.\\

\figcomment{\begin{figure}
\plotone{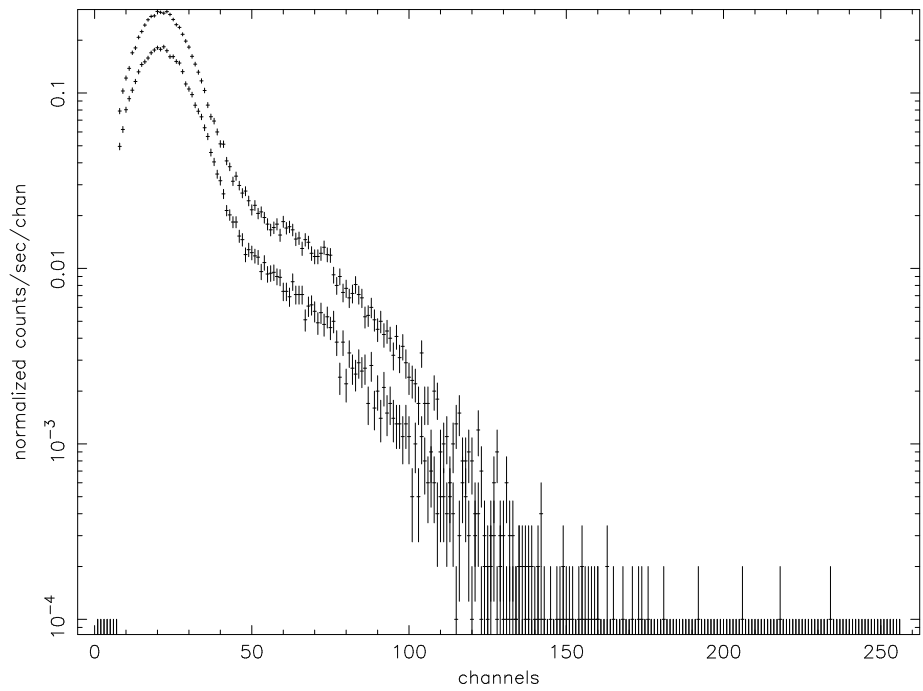}
\caption{The calculated model spectra convolved with the ROSAT PSPC response
matrix for $\varphi=0$ (upper points) and $\varphi=\pi/2$ (lower
points).  The model parameters are described in the text.  The
errorbars are for an exposure of $10^4$ s.}
\label{fig:xspec}
\end{figure}}
\end{appendix}

\figcomment{
\end{document}